\documentclass[pra,10pt,twocolumn,aps,floatfix,amsmath,amssymb,amsfonts,superscriptaddress]{revtex4-1}
\usepackage{amsmath,amssymb,dsfont,graphicx, float}
\usepackage{color}
\usepackage{dcolumn}
\usepackage[T1]{fontenc}

\begin{document}

\newcommand{\bra}[1]{\langle #1|}
\newcommand{\ket}[1]{| #1\rangle}
\newcommand{\tr}[1]{{\rm Tr}\left[ #1 \right]}
\newcommand{\av}[1]{\left\langle #1 \right\rangle}
\newcommand{\proj}[1]{\ket{#1}\bra{#1}}
\newcommand{\x}{{\bf r}}
\newcommand{\xp}{{\bf r}_\perp}
\newcommand{\bk}{\mathbf{k}}

\newcommand{\aho}{a_\perp}

\def\beq{\begin{equation}}
\def\eeq{\end{equation}}

\newcommand{\B}[1]{\mathbf{#1}}
\newcommand{\eq}[1]{Eq.~\eqref{#1}}
\newcommand{\floor}[1]{\lfloor #1 \rfloor}\
\newcommand{\zi}[1]{\textcolor{red}{#1}}
\newcommand{\ms}[1]{\textcolor{blue}{#1}}

\title{Analytically solvable quasi-one-dimensional Kronig-Penney model}

\author{Marta Sroczy\'{n}ska}
\affiliation{Faculty of Physics, University of Warsaw, ul. Pasteura 5, PL--02--093 Warszawa, Poland}

\author{Tomasz Wasak}
\affiliation{Faculty of Physics, University of Warsaw, ul. Pasteura 5, PL--02--093 Warszawa, Poland}
\affiliation{Max Planck Institute for the Physics of Complex Systems, N\"othnitzer~Str.~38, 01187 Dresden, Germany}

\author{Zbigniew Idziaszek}
\affiliation{Faculty of Physics, University of Warsaw, ul. Pasteura 5, PL--02--093 Warszawa, Poland}

\begin{abstract}
We generalize the textbook Kronig-Penney model to realistic conditions for a quantum-particle moving in the quasi-one-dimensional (quasi-1D) waveguide, where motion in
the transverse direction is confined by a harmonic trapping potential. Along the waveguide, the particle scatters on an infinite array of regularized delta
potentials. Our starting point is the Lippmann-Schwinger equation, which for quasi-1D geometry can be solved exactly, based on the analytical formula for the quasi-1D
Green's function. We study the properties of eigen-energies as a function of particle quasi-momentum, which form band structure, as in standard Kronig-Penney model. We test
our model by comparing it to the numerical calculations for an atom scattering on an infinite chain of ions in quasi-1D geometry. The agreement is fairly good and can be
further improved by introducing energy-dependent scattering length in the regularized delta potential. The energy spectrum exhibits the presence of multiple overlapping
bands resulting from excitations in the transverse direction.  At large lattice constants, our model reduces to standard Kronig-Penney result with one-dimensional
coupling constant for quasi-1D scattering, exhibiting confinement-induced resonances. In the opposite limit, when lattice constant becomes comparable to harmonic
oscillator length of the transverse potential, we calculate the correction to the quasi-1D coupling constant due to the quantum interference between scatterers. Finally,
we calculate the effective mass for the lowest band and show that it becomes negative for large and positive scattering lengths.
\end{abstract}

\maketitle

\section{Introduction}

The standard textbook Kronig-Penney (KP) model~\cite{Kronig1931}, introduced in courses on solid states physics~\cite{Kittel2005}, is defined in one dimension (1D) and
provides probably the most simple picture of a crystalline solid. In this approach electron-ion interactions are replaced by rectangular wells or Dirac $\delta$-potentials,
allowing for analytical solution of the problem, and derivation of the energy spectrum exhibiting band structure behavior.  Despite its educational values, there are
numbers of laboratory systems for which KP model can be applied, at least for qualitative description. 

In solid states, periodic one-dimensional structures of atoms can be created using scanning tunneling microscopy technique~\cite{Nilius2002,Ortega2007,Oncel2008}. In
experiments with ultracold atoms, periodic one-dimensional structures of atoms can be realized in optical lattices in the Mott insulator phase~\cite{Bloch2005}. Using two
distinct species of atoms and applying the technique of {\it species-selective dipole potential}, i.e. an optical potential experienced exclusively by one species
\cite{LeBlanc2007,Catani2009}, in principle one could realize situation resembling conditions of KP model. In such a realization one species of atoms is tightly confined in
the optical lattice potential, creating an effective periodic lattice for the atoms of the second species, which do not feel the optical lattice potential.  Hybrid
systems of ultracold atoms and ions~\cite{Harter2014,Tomza2019} provide another realization of KP model, where 1D crystal of ions can be combined with ultracold atomic
cloud in quasi-1D geometry. Such a system offers a powerful platform for quantum simulations of a solid state including inherently electron-phonon
coupling~\cite{Bissbort2013}. Here, the atom-ion interaction results in even and odd scattered waves in 1D geometry~\cite{Idziaszek2007controlled}, which leads to
a generalized KP model including two separate coupling constants for even and odd partial waves~\cite{Negretti2014}.

While the optical lattices potentials are typically far from the assumption of KP model since the characteristic size of barriers is limited by the wavelength of light,
in recent experiment subwavelength optical structures were obtained~\cite{Lacki2016,Wang2018}, making it possible to directly trap the atoms in optical lattices of nearly
$\delta$-potentials. The common assumption for all these realizations is that motion of atoms in the transverse direction can be neglected due to the strong confinement,
making these systems effectively 1D. Nevertheless, at small distances when atoms approach the scattering center (atom of a second species, ion, or a subwavelength optical
barrier) the scattering process takes place in three dimensions, and as such requires inclusion of the transverse motion, which is neglected by definition in the standard
KP model.

In our work, we generalize KP model to include the motion in the transverse direction, which allows for a realistic description of the scattering process between the
particle and scattering centers (impurities) at the lattice nodes.  We consider the motion of an atom in quasi-1D geometry in the presence of periodically spaced
impurities, either atoms or ions.  The impurities are distributed at equidistant points along $z$ at the center of the trap. They are interacting with the moving atom,
thus, forming an external periodic potential acting on that particle. In this work, we assume that the mobility of impurities is negligible. The atom can move freely in
the $z$ direction, whereas the motion in the perpendicular directions is confined by the harmonic trapping potential, $V_\perp(\xp) = \frac{1}{2}m\omega^2 \xp^2$, where
by $\xp=(x,y)$ we denote the coordinates perpendicular to the $z$-axis. In the case of atomic impurities, at ultracold temperatures the atom-atom scattering can be
accurately modeled by Fermi zero-range pseudopotential~\cite{Fermi1936,Huang1987} parametrized by the $s$-wave scattering length. In the case of ionic lattice, we employ
directly in the calculations the long-range atom-ion polarization potential characterized by the characteristic range of the potential $R^\ast$
\cite{Idziaszek2009,Tomza2019}.

Solving the Schr\"odinger equation either with the Green's function method or numerically using the finite element method, we determine the wavefunctions and spectrum of
stationary states. Inclusion of the transverse degrees of freedom permits us to study the interplay between the transverse confinement length and remaining parameters:
the scattering length, the characteristic range of the potential, and a lattice constant, which is beyond the scope of the standard KP model. 

In the limit of large impurity separations, one can expect that atom-impurity scattering can be described as a single separate scattering process. In such a case the
scattering exhibits confinement-induced resonances and the system properties can be obtained from the usual KP model with 1D coupling constant calculated for quasi-1D
geometry derived by Olshanii \cite{Olshanii1998,Bergeman2003}. In the opposite limit, when lattice spacing is comparable or smaller to the transverse confinement length
one expects that scattering processes on neighboring impurities will interfere, and in such a case quasi-1D KP model cannot be reduced to purely 1D KP model.

The paper is structured as follows. In Section II we discuss physical systems where quasi-1D KP model can be realized and we specify the interaction potentials between
atoms and impurities belonging to the lattice. In Section III we demonstrate how to derive the ordinary KP model using the Green's function method. The same technique is later
applied to solve analytically the quasi-1D KP model, obtaining equations determining eigenergies as a function of quasi-momentum. Section V presents details of our
numerical calculations performed for the model system containing an atom moving in quasi-1D trap and interacting with the infinite lattice of ions. The analytical results
derived in Section IV for regularized delta potentials are compared in Section VI to the numerical calculations for the model hybrid atom-ion system in quasi-1D
geometry. We analyze the properties of the energy spectrum, the wave functions, and effective mass on the atom-ion scattering length. Section VII presents
conclusions, while four appendices present some technical details on analytical calculations for quasi-1D KP model.




\section{Physical system}
In this work we consider motion of an atom in quasi-1D geometry in the presence of a periodic lattice of impurities. The Schr\"odinger equation for the atom of mass $m$ moving in the presence of external trap and lattice of impurities reads
\begin{equation}
\label{schro}
-\frac{\hbar^2 \nabla^2}{2m} \psi(\x) + V_\perp(\xp)\psi(\x) + V(\x) \psi(x) = E \psi(x),
\end{equation}
where $V(\x)$ is the periodic potential, with period $L$ in $z$-direction (taken along unit vector $\mathbf{e}_z$),
\begin{equation}
\label{latt}
V(\x)=V(\x+ L \mathbf{e}_z) = \sum_{n=-\infty}^{\infty} V_{\mathrm{ai}}(\x-{\bf d}_n), 
\end{equation}
given in terms of a sum of individual atom-impurity potentials $V_{ai}(\x)$. In this notation, the vector $\mathbf{d}_n=(0,0,nL)$ is the position of the $n$-th impurity
on the $z$-axis. The transverse confinement potential $V_\perp(\xp)$ is simply
\begin{equation}
V_\perp(\xp) = \frac{1}{2}m\omega^2 \xp^2
\end{equation}
with $\xp=(x,y)$. 

\subsection{Potential for atom-atom scattering}

In the case of atomic impurities in the ultracold regime, the scattering takes place dominantly in $s$-wave, and the atom-impurity collisions can be accurately 
modeled by Fermi zero-range pseudopotential~\cite{Fermi1936,Huang1987}, 
\beq
V_{ai}(r) = g \delta(\x) \frac{\partial}{\partial r} r,
\label{eqn:Fermi}
\eeq
where the coupling strength $g = 2\pi \hbar^2 a / m$ is related to the $s$-wave scattering length $a$ and the mass of the atom $m$. Note that the mass of the impurity does not enter the considerations, because of the assumption of the immobility of the impurity; formally it is equivalent to taking the limit of infinite impurity's mass in which case the reduced mass is equal to $m$.

The accuracy of Fermi pseudopotential can be further improved by introducing an energy-dependent pseudopotential \cite{Blume2002fermi,Julienne2002effective}
\beq
V_{ai}(r;E) = \frac{2\pi \hbar^2 a(E)}{m} \delta(\x) \frac{\partial}{\partial r} r,
\eeq
which is defined in terms of an energy-dependent scattering length
\beq
a(E) = - \frac{\tan \delta_0(k)}{k},
\label{aeff}
\eeq
where $\delta_0(k)$ is the $s$-wave phase shift and wave-vector $k$ is related to the kinetic energy $E = \hbar^2 k^2/(2m)$. Such a description extends the validity of $\delta$-pseudopotential to finite kinetic energies and to the finite range of the interaction potential, where characteristic range becomes comparable to the external confinement.     

\subsection{Potential for atom-ion scattering}

The long-range part of the atom-ion interaction potential reads
\beq
V_{ai}(r)  \xrightarrow[]{r\rightarrow \infty} -\frac{C_{4}}{r^{4}},
\label{Vai}
\eeq
where the constant $C_4$ depends on the electric dipole polarizability $\alpha$ of an atom in the electronic ground state $C_4 = \alpha e^2/(8\pi\epsilon_0)$.
Such a potential is valid provided that an atom is in its electronic ground state. Otherwise, if it is in an excited state, it has nonzero quadrupole moment, and the charge-quadrupole dominates the long range part of the atom-ion interaction potential \cite{Mies1973}.

The introduced interaction potential defines the length scale $R^*~=~\sqrt{2mC_4/\hbar^2}$ and energy scale $E^*~=~\hbar^2/(2m(R^*)^2)$. Note that these units are defined
in a slightly different way than units usually defined in the context of atom-ion collisions \cite{Idziaszek2009,Tomza2019}, where a reduced mass of the atom-ion system
usually enters in the problem, instead of $m$.

At short distances realistic ion-atom interaction appears to be strongly repulsive and the singular potential \eqref{Vai} has to be regularized. One of the possible ways
is to introduce a cut-off radius $b$, so that the potential becomes \beq V_{\mathrm{ai}}(r) = -\frac{C_4}{(r^2 + b^2)^2}, \label{eqn:potRegB} \eeq and the relation
between $b$ and scattering length is given by \cite{szmytkowski} \beq
a(b)=R^*\sqrt{1+\left(\frac{b}{R^*}\right)^2}\cot\left(\frac{\pi}{2}\sqrt{1+\left(\frac{R^*}{b}\right)^2}\right).  \eeq One value of the scattering length can be
reproduced by many different values of $b$. The choice of $b$ in a given range determines how many bound states it will give. The potential supports $n$ bound states for
$b\in (b_{n-1}, b_{n})$, and  $b_n = 1/\sqrt{4n^2-1}$.

In our calculations we will also employ another approach based on the quantum-defect method \cite{Seaton1983,Greene1982,Mies1984}, where 
one introduces additional quantum-defect parameter, which is the phase $\varphi$, characterizing the short-range behavior of the wave function
for $R_0 \ll r \ll R^*$  \cite{Idziaszek2007controlled,Idziaszek2011,Gao2010}:
\beq
\psi(r)\sim \sin\left(\frac{R^*}{r}+\varphi\right).
\label{QuantDef}
\eeq

\section{1D Kronig-Penney model}
Before we start to solve Eq.~\eqref{schro} with Green's function method, we first demonstrate how this method works in the case of a purely 1D system. We
expect to reproduce the well-known solution of the strictly 1D KP model. The 1D situation is simpler to handle mathematically because no regularization of the
pseudopotential is required. The Sch\"odinger equation is then
\begin{equation}
\label{schro1D}
-\frac{\hbar^2}{2m}\frac{\partial}{\partial z} \psi(z) - E \psi(z) = - V_{1D}(z) \psi(z),
\end{equation}
where $V_{1D}(z) = \sum_n g_{1D}\delta(z-nL)$ and the summation extends over all integer values. The coupling constant $g_{1D}$ is the given parameter of the problem.
This equation can be formally solved by the one-dimensional Green's function $G_{1D}(z,z|E)$ defined by Eq.~\eqref{green_def}.
The formal solution is given by
\begin{equation}
\label{1dsol}
\psi(z) = \mathcal{C}\psi_0(z) + \int\!\!dz'\, G_{1D}(z,z'|E)V_{1D}(z') \psi(z'),
\end{equation}
where $\psi_0(z)$ is the solution of Eq.~\eqref{schro1D} with $V_{1D}\equiv0$. Inserting here the potential, we obtain 
\begin{equation}
\label{eq1D}
\psi(z) = \mathcal{C}\psi_0(z) + \sum_n g_{1D} G_{1D}(z,d_n|E) \psi(d_n),
\end{equation}
where $d_n = n L$ is the position of the impurity. 

Before we go proceed further, we invoke the Bloch theorem, which states that the wavefunction of the hamiltonian with discrete
translational symmetry has to be also an eigenfunction of the translation symmetry operator. It means that $\psi(z) = e^{i q z }u_q(z)$ with periodic $u_q(z+L)=u_q(z)$. 
Consequently, $\psi(z+L) = e^{i\theta}\psi(z)$ and $\psi(z+n L) = e^{in\theta}\psi(z)$ with the phase $\theta = qL$. 
We use this property to solve Eq.~\eqref{eq1D}. To this end, we first set $z=d_{n'}$.
Next, we have the freedom to choose the Green's function; we take $G_{1D}^+$ which describes the outgoing scattered wave. It is straightforward to show, that any linear combination
$G_p=p G_{1D}^++(1-p) G_{1D}^-$ leads to the same result provided $0\leqslant p \leqslant 1$ (this condition is necessary for $G_p$ to be a Green's function).

In the considered case, the Eq.~\eqref{eq1D} leads to 
\begin{equation}
\label{1dfin}
e^{i\theta n'} = \mathcal{C}\psi_0(d_{n'}) + \sum_n g_{1D} G_{1D}^+(d_{n'},d_n|E) e^{i\theta n},
\end{equation}
in which, without loss of generality, we set $\psi(d_n) = e^{i \theta n}$, and the angle $\theta = q L$ satisfies, according to Bloch's theorem, $-\pi\leqslant \theta \leqslant \pi$.
This condition for $\theta$ ensures that $q$ is taken from first Brillouin zone.

Assuming now $E=\hbar^2k^2/2m\geqslant0$ and inserting into Eq.~\eqref{1dfin} an explicit formula for the Green's function (Eq.~\eqref{G1D} in appendix \ref{Sec:AppG1D}), we obtain
\begin{equation}
1- \frac{m g_{1D}}{\hbar^2}\frac{1}{i k} \sum_n e^{ik L |n|} e^{i \theta n} = \mathcal{C}\psi_0(d_{n'})e^{-i\theta n'}.
\end{equation}
The right hand side is a function of $n'$, whereas the left-hand side is independent of $n'$. Consequently, we set $\mathcal{C}=0$ and arrive at
\begin{equation}
\label{KP1}
\cos{\theta}= \cos{k L} + \frac{m g_{1D}}{\hbar^2} \frac{\sin{ k L}}{k},
\end{equation}
which is the well-known relation describing the dispersion relation $E(k)=\hbar^2 k^2/2m$ as a function of quasi-momentum $q=\theta L^{-1}$ in the Kronig-Penney model. An analogous calculation
for $E= - \hbar^2 \kappa^2/2m<0$ leads to Eq.~\eqref{KP1} with $k$ interchanged by $i\kappa$, which is related to analytic continuation of the wavefunction from $E>0$ to $E<0$ passing
by $E=0$ on the physical sheet, i.e., in the upper plane of the complex energy $E$.

\section{Quasi-1D Kronig-Penney model}

Having established the technique of the Green's function to solve one-dimensional KP model, we now proceed to solve the quasi-1D case.
In analogy to \eq{1dsol}, the formal solution of \eq{schro} with the potential \eq{latt} is given by the 3D Green's function:
\begin{equation}
\label{3dsol}
\psi(\x) = \int\!\! d^3r'\, G_{3D}(\x,\x'|E)V(\x') \psi(\x'),
\end{equation}
where we dropped the homogenous term. The explicit form of the function $G_{3D}$ is presented in Appendix \ref{G3Ddef}. Inserting here the form of the potential we are led to
\begin{equation}
\label{3dsolg}
\psi(\x) = \sum_{n=-\infty}^{+\infty}  G_{3D}(\x,\B{d}_n |E) g  \gamma_n  ,
\end{equation}
with 
\begin{equation}
\label{defgamma}
\gamma_n = \frac{\partial}{\partial r_n}\bigg(r_n \psi(\x') \bigg)\bigg|_{\x \to \B{d}_n}.
\end{equation}
This equation for the wavefunction resembles the one given by \eq{eq1D}. However, they are different in one important aspect. Due to the 3D interaction with the impurity,
the 3D wavefunction $\psi(\x)$ when $\x$ is in the neighbourhood of the $n$-th impurity behaves as $\propto 1 - a/|\x - \B d_n|$ in the ultracold regime where the
$s$-wave scattering dominates. Consequently, in our contact potential approximation, the wavefunction has a pole at $\x=\B{d}_n$ and the value of the waverfunction cannot
appear in \eq{3dsolg} in the same manner as in \eq{eq1D}. Nevertheless, the derivative in the form of $\gamma_n \propto \partial( r\psi)/\partial r$ is finite at $\x=\B
d_n$ and is related to the atom-impurity scattering length $a$.

To go further, we invoke the Bloch theorem. To this end, we note that $\gamma_n$ and $\gamma_{n+1}$ are related to the derivatives of $\psi(\x)$ calculated around $\B
d_n$ and $\B d_{n+1}$. Since these points are separated by a lattice constant $L$, they differ by a phase $e^{i \theta}$ with $\theta = q L$. Without loss of
generality, we may take $\gamma_n = e^{ i n \theta }$. To find the relation between $\theta $ and $E$ we insert \eq{3dsolg} into \eq{defgamma}. The resulting
expression includes the terms  $G_{3D}(\B d_n,\B d_{n'}|E)$ with $n\neq n'$, because the regularizing operator acts only when $\x \approx \B d_{n'} $ in $G_{3D}(\x,\B
d_{n'}|E)$. Consequently, the derivative in \eq{defgamma} modifies only diagonal terms of the Green's function leaving the off-diagonal unaffected. The resulting expression
takes the following form:
\begin{equation}
\label{eq3Da}
\gamma_n = g\bigg( \gamma_n \beta(E) + \sum_{n'\neq n} \gamma_{n'} G_{3D}(\B d_{n'}, \B d_n |E) \bigg),
\end{equation}
where 
\begin{equation}
\label{betadef1}
\beta(E) = \frac{\partial }{\partial r}\bigg( r G_{3D}(\x,0|E)  \bigg)\bigg|_{\x=0}.
\end{equation}
Notice, that due to the translation symmetry along $z$-axis, the Green's function $G(x,y,z;y',y',z'|E)$ depends only on the distance $|z-z'|$. Moreover, since all the impurities are
residing only on the $z$-axis, we may take $\xp=\xp'=0$. Therefore, hereafter by writing $G_{3D}(z-z'|E)$ we refer to $G_{3D}(0,0,z;0,0,z'|E)$. Using this new notation, and
exploiting the result $\gamma_n = e^{i n \theta }$, we arrive at
\begin{equation}
\label{eq3Db}
1 = g\bigg(  \beta(E) + \sum_{n'>0}  2  G_{3D}( n' L |E)\, \cos{n'\theta} \bigg),
\end{equation}
where now
\begin{equation}
\label{betadef2}
\beta(E) = \frac{\partial}{\partial z}\bigg(z G_{3D}(z|E)\bigg)\bigg|_{z=0^+}.
\end{equation}

To evaluate all the necessary quantities we take the ``Feynman Green's function'' $G_{3D}^F = (G_{3D}^+ + G_{3D}^-)/2$.
The detailed calculations of $\beta(E)$ and the sum with $n'>0$ are given in Appendices \ref{Appbeta} and \ref{AppLambda}. The resulting expression for $\beta(E)$ is given by
\begin{equation}
\label{beta_final}
\beta(E) = - \frac{1}{\pi a_\perp^2} \frac{2m}{\hbar^2} \frac{a_\perp}{4} \zeta_H\bigg( \frac12, 1 - \frac{E-E_{2n^*}}{2\hbar \omega} \bigg),
\end{equation}
where we introduced the energy-dependent nearest threshold energy $E_{2n^*} = \hbar\omega(2 n^*(E) + 1)$, and $\zeta_H$ is the Hurwitz zeta function.  The energy
$E_{2n^*}$ plays in the problem an important role. It is the energy corresponding to the $n^*$-th 2D harmonic oscillator state with $m=0$. Note that $E_{2n^*} $ depends on $E$ 
and that $E_{2n^*}$ is always below $E$, but increasing $n^*$ by one takes $E$ below this new value, i.e., $E_{2n^*} \leqslant E < E_{2(n^*+1)}$. Alternatively, we may write
$n^*(E) = \floor{(E-\hbar\omega)/2\hbar\omega}$, where $\floor{x}$ is the floor function.

The sum over $n'$ in \eq{eq3Db} is conveniently split into two terms, because they differ in asymptotics if $L \gg a_\perp$. Therefore, we define
\begin{subequations}
	\label{defLambda}
	\begin{eqnarray}
	\Lambda(E) &\equiv& \sum_{n>0}  2  G_{3D}( n L |E)\, \cos{n\theta} \\
	&=& \frac{2m}{\hbar^2}\frac{L}{\pi a_\perp^2}\bigg(\Lambda_p(E) + \Lambda_e(E)\bigg),
	\end{eqnarray}
\end{subequations}
with
\begin{equation}
\label{eqn:LambdaP}
\Lambda_p(E) =  \sum_{0 \leqslant n \leqslant n^*} \frac{1}{2k_n L} \frac{\sin{k_n L}}{\cos\theta - \cos{k_n L}},
\end{equation}
with $k_n a_\perp = 2\sqrt{ (E-\hbar\omega)/2\hbar\omega - n}$
and 
\begin{equation}
\Lambda_e(E) =   \sum_{n \geqslant 1} \frac{1}{k_n L} \mathrm{Re}\bigg[\frac{1}{1 - e^{\tilde{k}_n L + i \theta}} \bigg],
\label{eqn:LambdaE}
\end{equation}
with $\tilde k_n a_\perp = 2\sqrt{n - (E-E_{2n^*})/2\hbar\omega}$. The equation determining energy as a function of $\theta$ is given by
\begin{equation}
\label{final_eq}
1 =  \frac{a}{a_\perp} C(E) + \frac{4 a L}{a_\perp^2}(\Lambda_p(E) + \Lambda_e(E)),
\end{equation}
where $C(E) = - \zeta_H(1/2, 1 - {(E-E_{2n^*})}/{2\hbar \omega})$. This can be further transformed to the following form
\begin{equation}
\label{final_eq1}
a_{1D}(E) =  - 2 L (\Lambda_p(E) + \Lambda_e(E)),
\end{equation}
where $a_{1D}(E)$ is the one-dimensional energy-dependent scattering length in quasi-1D system
\begin{equation}
a_{1D}(E) = - \frac{a_\perp^2}{2 a}\bigg( 1 - C(E) \frac{a}{a_\perp} \bigg).
\label{eqn:a1DOlshaniiE}
\end{equation}
It is a generalization of the 1D scattering length to the case of finite energies \cite{Bergeman2003,Naidon2007}. In the low-energy limit, $E \to \hbar \omega$, it reduces to the well-known result, 
\begin{equation}
a_{1D} = - \frac{a_\perp^2}{2 a}\bigg( 1 - C \frac{a}{a_\perp} \bigg).\label{eqn:a1DOlshanii}
\end{equation}
with $C = C(\hbar\omega)= 1.46035\ldots$. which was first derived in the seminal paper by Olshanii~\cite{Olshanii1998}.

The quantities above are so defined that $k_n$ and $\tilde k_n$ are always real and positive. $\Lambda_p(E)$ results from all the physically open channels of the 2D radial
harmonic oscillator. For fixed energy $E$ only radial excitations with energies up to $E_{2n^*} \leqslant E$ contribute to this term.  If energy $E$ is slightly above
$\hbar\omega$ only a single channel is open, and all the others are closed. This is why only periodic functions appear in $\Lambda_p$ (and so the subscript $p$). On the other hand, the term $\Lambda_e(E)$ contains exponential function which is related to the fact that it captures the contribution from the closed channels with radial
excitations with energies larger than $E$.

Let us now see how, in the limiting case, we reproduce 1D KP model. To this end, we assume that the energy $E= \hbar \omega + \hbar^2 k^2 /2m$ is slightly above the
first threshold $E_0 = \hbar\omega$ but far from the second threshold $E_2 = (2 \times 1 + 1)\hbar \omega = 3\hbar\omega$. This means that the kinetic energy $\hbar^2 k^2 /2m \ll \hbar \omega$, $n^*=0$ and $C(E)$ can be approximated by its low-energy limit 
$C(E) \approx C$.
Next, we notice that $k_0 = k$ and $\Lambda_p(E)$ boils down to a single term only:
\begin{equation}
\Lambda_p(E) \approx \frac{1}{2 k L} \frac{\sin k L}{\cos\theta - \cos k L}.
\end{equation}
Now, the function $\Lambda_e$ in the denominator has exponents of $\tilde k_n L  = (2L / a_\perp)\sqrt{n - (k L/2)^2}$. If the momentum $k$ is far from the first threshold, then in the limit $L \gg a_\perp$ the exponent is large. Consequently, $\Lambda_e$ is exponentially small, and can be neglected. In such a case, Eq.~\eqref{eqn:a1DOlshaniiE} reduces to \eqref{KP1}  determining the energy spectrum in the 1D Kronig-Penney model with the one-dimensional coupling constant $g_{1D} = -\hbar^2/m a_{1D}$ given in terms of the one-dimensional scattering length $a_{1D}$~\eqref{eqn:a1DOlshanii}. This situation corresponds physically to the scattering on the well-separated impurities, where each of the scattering events is described by the quasi-1D result of Olshanii \cite{Olshanii1998}. 

In the limit, when distance between impurities is comparable to the transverse scattering length, or smaller: $L \lesssim a_{\perp}$, the contribution from the term $\Lambda_e$ has to be included. Assuming again that the energy is slightly above the first threshold: $\hbar^2 k^2 /2m \ll \hbar \omega$ and $n^*=0$, we have $\tilde k_n a_\perp \approx 2\sqrt{n}$, which after substituting into \eqref{eqn:LambdaE} gives 
\begin{equation}
\Lambda_e(\theta) \approx   \sum_{n \geqslant 1} \frac{a_\perp}{2 \sqrt{n} L} \mathrm{Re}\bigg[\frac{1}{1 - e^{2 \sqrt{n}L/a_\perp + i \theta}} \bigg],
\label{eqn:LambdaEappr}
\end{equation}
This can be further simplified to
\beq
\Lambda_e (\theta)\approx-\frac{1}{2}\frac{a_{\perp}}{L}\cos(qL)H\left(2 \frac{L}{a_{\perp}}\right),\label{eqn:LambdaEapprWithH}
\eeq 
where 
\beq
H(x) = \sum_{n=1}^{\infty} n^{-1/2}\exp\left(-\sqrt{n} x \right). \label{h}
\eeq  
Within this regime Eq.~\eqref{eqn:a1DOlshaniiE} determining energy levels can be expressed in the following form:
\begin{equation}
a_{\mathrm{1D}}^{\mathrm{eff}}  =-  \frac{1}{k} \frac{\sin k L}{\cos\theta - \cos k L},
\end{equation}
that is equivalent to Eq.~\eqref{KP1} of 1D K-P model, with an effective one-dimensional scattering length
\begin{equation}
a_{\mathrm{1D}}^{\mathrm{eff}}(\theta)= a_{\mathrm{1D}}+2 L\Lambda_e (\theta).
\label{eqn:a1Deff}
\end{equation}
This quantity already incorporates the interference effects from the scattering on closely located impurities. For well-separated impurities ($L \to \infty$), $L\Lambda_e (\theta) \to 0$, and $a_{\mathrm{1D}}^{\mathrm{eff}}(\theta)$ reduces to one-dimensional scattering length $a_{\mathrm{1D}}$ derived for a single impurity in quasi-1D geometry. We note that this quantity depends on the quasi-momentum $q$, while we neglected the dependence of $a_{1D}$ on $k$.

\section{Atom moving in ionic lattice}

In this section we consider an atom confined in quasi-1D geometry and moving in the potential of a periodic infinite chain of ions. Exploiting the fact that the system is axially symmetric and periodic, according to the Bloch theorem, we can write the solution to the Schr\"odinger Eq.~\eqref{schro} in cylindrical coordinates $\rho$~and~$z$
\beq
\psi(\textbf{r}) = \frac{\exp(iqz)u_q(\rho, z)}{\sqrt{\rho}}\label{eqn:PsiFk}.
\eeq
Substituting \eqref{eqn:PsiFk} into the Schr\"{o}dinger equation \eqref{schro} with the periodic potential \eqref{latt} leads to the following equation for $u_q$ 
\beq
\begin{split}
-\frac{\hbar^2}{2m}\left(q^2 + 2iq\frac{\partial}{\partial z} + \frac{\partial^2}{\partial z^2} + \frac{\partial^2}{\partial \rho^2}  \right) u_q(\rho, z)+\\ + \frac{1}{2} \left(m\omega^2\rho^2 - \frac{\hbar^2}{4m}\frac{1}{\rho^2}\right)u_q(\rho, z)+\\
-\sum_{n=-\infty}^{\infty}V_{\mathrm{ai}}(r_n)u_q(\rho, z) = E u_q(\rho, z). 
\label{eqn:SchroedU}
\end{split}
\eeq

\subsection{Numerical calculations}

\begin{figure}[H]
	\includegraphics[width=0.48\textwidth]{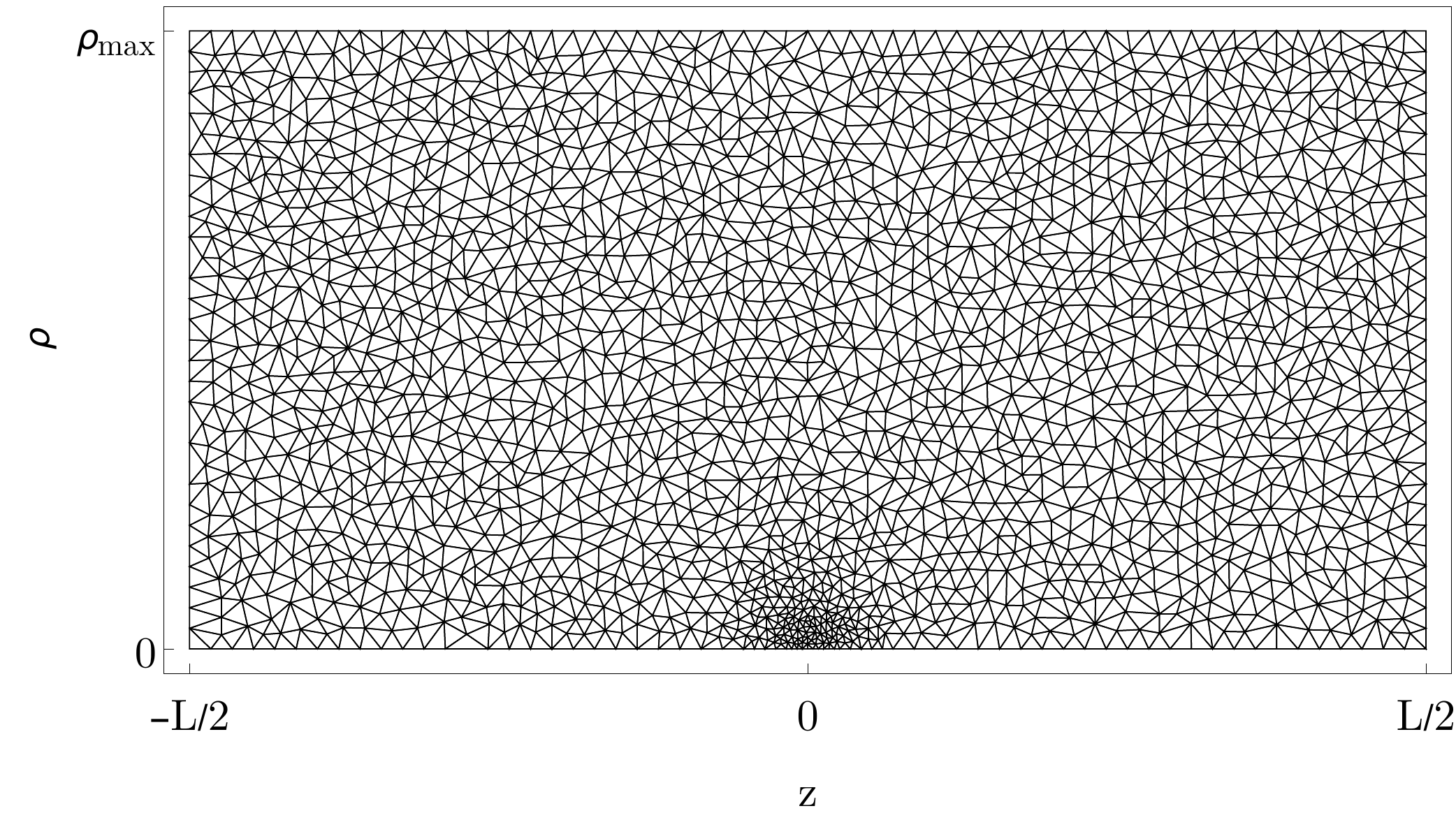}

	\caption{An example grid used for the finite element method. The grid size is determined by the local de Broglie wavelength and it becomes very dense close to the location of ion at $z=0$ and  $\rho=0$}
	\label{fig:mesh}
\end{figure}

In order to find $u_q$ for a given $q$, we solve numerically Eq.~\eqref{eqn:SchroedU} on a rectangle of size $L$ along $z$ coordinate and $\rho_{\mathrm{max}}$ along
$\rho$ coordinate (see Fig.~\ref{fig:mesh}).  We assume periodic boundary conditions in $z$: $u_q(\rho, -L/2) = u_q(\rho, L/2)$, Dirichlet boundary condition in
$\rho_{\mathrm{max}}$: $u_q(\rho_{\mathrm{max}}, z) = 0$ and von Neumann boundary condition at $\rho=0$: $\frac{\partial}{\partial\rho} \frac{u_q(\rho, z)}{\sqrt{\rho}} =
0$. The ion is placed at $z=0$ and $\rho=0$, however, the interaction potential $V(\x)$ contains contributions from the whole ion chain. In the numerical calculations we
apply the regularized form of the atom-ion interaction given by \eqref{eqn:potRegB}. The value of cut-off radius $b$ was adjusted to reproduce a given value of the $s$-wave
scattering length. In the numerical calculations we assumed $\rho_{\mathrm{max}}/a_{\perp} = 5$. We also chose the value of the parameter $b$ such that it supports only
one bound state as it gives relatively shallow potential, which is more convenient for numerical calculations.

The calculations are done using the finite element method routines built in Wolfram Mathematica program \cite{Mathematica}. Due to the fact that potential
\eqref{eqn:potRegB} becomes relatively deep in the vicinity of the ion's position, and the corresponding wave function is quickly oscillating in that region, in our
calculations we have used variable grid size. The mesh refinement function was related to the local de Broglie wavelength $\lambda(\x,E)=2\pi/\sqrt{2m|E -
  V_{\mathrm{ai}}(\rho, z)|/\hbar^2}$), by assuming that area of a single cell in the grid fulfils $\Delta \leq \lambda(\x,E)^2/N^2$. We have tested several values of $N$
parameter, observing that numerical calculations converge for $N \gtrsim 20$, and for the final results we assumed $N = 22$.  An example grid is shown in
Fig.~\ref{fig:mesh}.

\subsection{Quasi-1D KP model with energy-dependent scattering length}

The regularized delta pseudopotential is applicable, provided that the size of external trapping potential is much larger than characteristic range: $a_\perp \gg
R^\ast$. However, the validity of the Fermi pseudopotential can be extended to the regime $a_\perp \sim R^\ast$ by replacing scattering length $a$ by the energy-dependent
scattering length $a(E)$ defined in Eq.~\eqref{aeff} \cite{Julienne2002effective,Blume2002fermi}. Consequently, the energy spectrum for the atom moving in quasi-1D geometry
and interacting with the infinite chain of ions can be obtained from Eq.~\eqref{final_eq1} with 1D energy-dependent scattering length
\begin{equation}
a_{1D}(E) = - \frac{a_\perp^2}{2 a(E)}\bigg( 1 - C(E) \frac{a(E)}{a_\perp} \bigg).
\label{eqn:a1DOlshaniiE1}
\end{equation}
Depending on the method of regularization at short distances, the energy-dependent scattering length $a(E)$  for the atom-ion interaction can be evaluated either numerically or analytically. For the regularization based on the cut-off radius, see Eq.~\eqref{eqn:potRegB}, the energy dependent scattering length was evaluated 
numerically by solving corresponding radial Schr\"odinger equation with Numerov method. In the framework of the quantum-defect theory, the atom-ion interaction is regularized by introducing some short-range quantum-defect parameter, eg. short-range phase, see Eq.~\eqref{QuantDef}, and the solution of the radial Schr\"odinger equation 
can be expressed in terms of Mathieu functions of the imaginary argument \cite{Vogt1954,Spector1964,Levy1963,Idziaszek2006}. In this way, one can 
find analytical expression for both $a(E)$ and its  low-energy expansion \cite{Idziaszek2011}
\beq
a(E)\approx a(0) + \frac{\pi}{3}(R^*)^2 k, \label{eqn:aE_approx}
\eeq
where $a(0) = \lim_{k\rightarrow0} a(k)$.

\begin{figure}[H]
\includegraphics[width=0.48\textwidth]{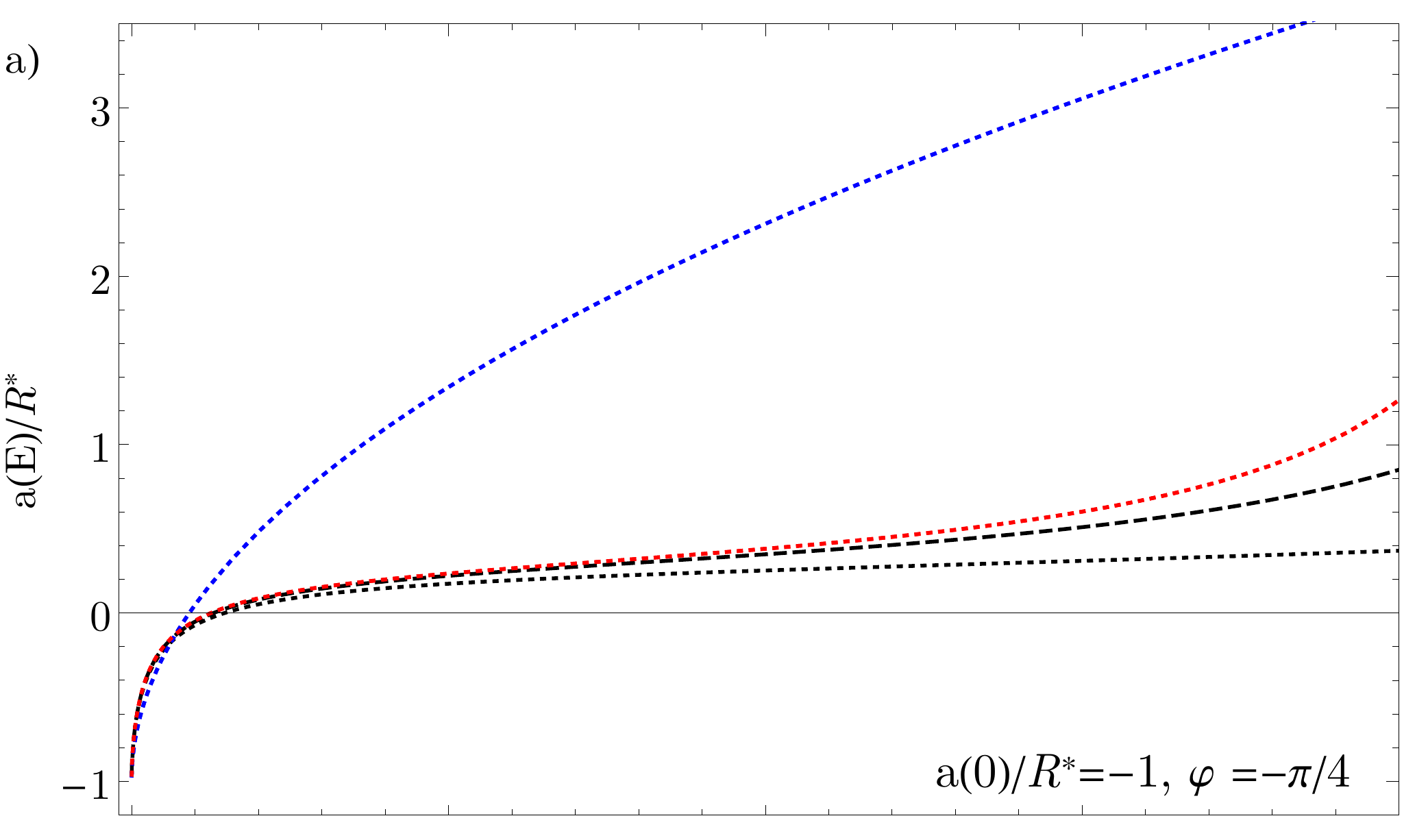}
\includegraphics[width=0.48\textwidth]{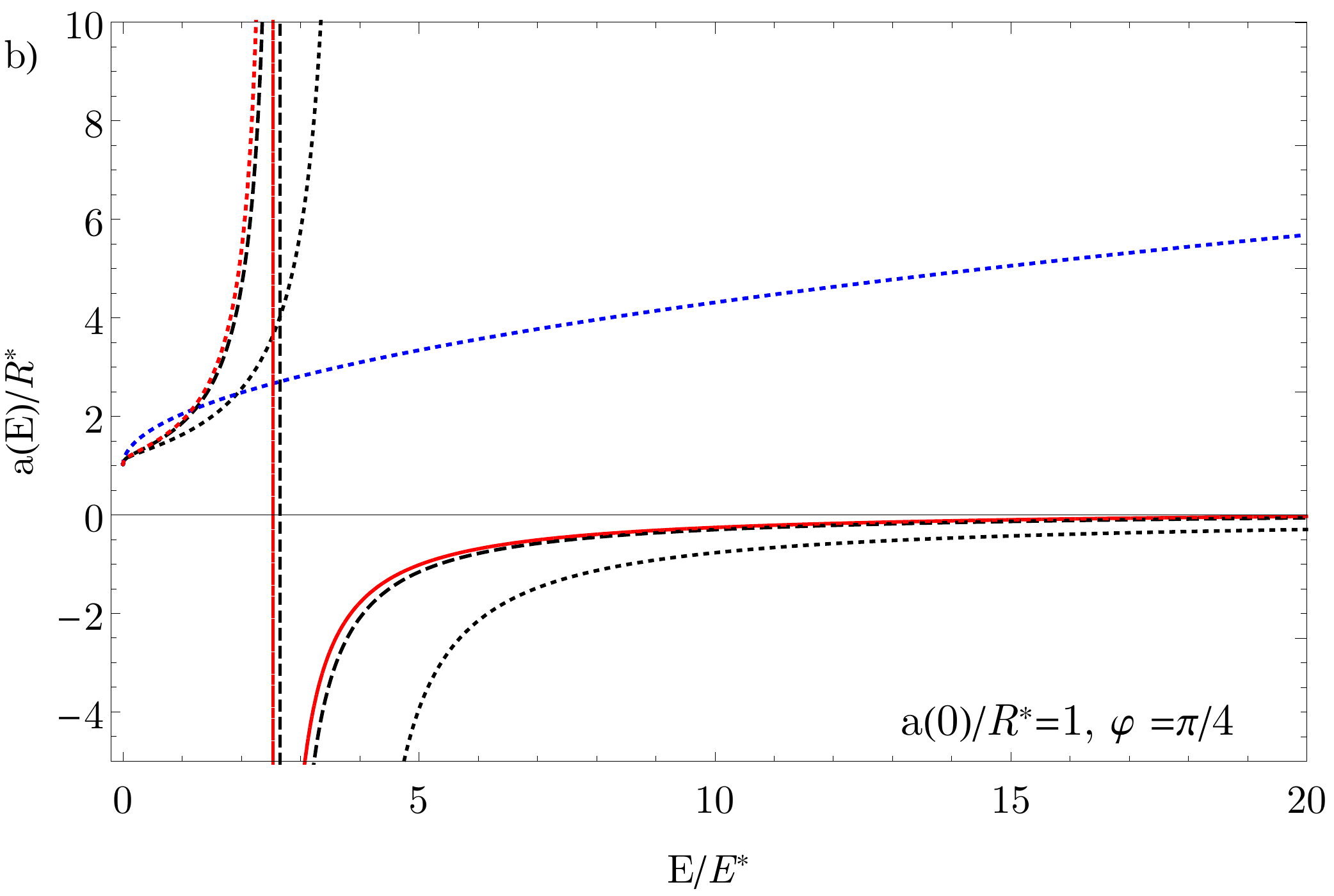}
	\caption{Energy-dependent $s$-wave scattering length calculated from Eq.~\eqref{aeff} (black dotted line corresponds to the value of $b$ supporting only one bound state, black dashed line corresponds to $b$ supporting three bound states) compared with the quantum-defect regularization parametrized by $\varphi$ (red dotted line) and  low-energy expansion Eq.~\eqref{eqn:aE_approx} (blue dotted line). The values of $a(0)$, the corresponding parameter $b$ and short-range phase $\varphi$ are: (a) $a(0)/R^*=-1$, $b/R^*=0.299$ (black dotted), $b/R^*=0.135$ (black dashed), $\varphi = \pi/4$, (b) $a(0)/R^*=1$, $b/R^*=0.431$ (black dotted), $b/R^*=0.156$ (black dashed), $\varphi = -\pi/4$. 
} \label{fig:aE}
\end{figure}

Fig. \ref{fig:aE} shows the energy-dependent scattering length for $a(0) = \pm R^\ast$. It compares numerical calculations for regularized potential \eqref{eqn:potRegB},
with the quantum-defect result [Eq.~(26) in Ref.~\cite{Idziaszek2011}] parametrized by $\varphi$ and the low-energy expansion \eqref{eqn:aE_approx}. In general, $a(E)$
calculated from the regularized potential \eqref{eqn:potRegB} agrees well with quantum-defect regularization, and the agreement improves with the number of bound states
supported by the regularized potential (smaller values of $b$). In that sense, quantum-defect approach represents the limit of $b \to 0$ of regularization
\eqref{eqn:potRegB}. On the other hand, low-energy expansion \eqref{eqn:aE_approx} is valid only for very small energies $E \ll E^\ast$, and it completely fails to
predict low-energy behaviour for $a(0) = R^\ast$, where the energy-dependent scattering length exhibits a resonance at $E \approx 2.53E^*$ (for the quantum-defect
method). Actually, position of resonances for the regularized potential \eqref{eqn:potRegB} slightly differ, and occur at $E=3.76E^*$ (in the case of potential supporting
one bound state) and at $E=2.63E^*$ (smaller $b$, potential supporting three bound states). Therefore, in the rest of the paper, the energy-dependent scattering length is
evaluated using the quantum-defect method.

\section{Results}\label{sec:numerical_results}

\subsection{Energy spectrum}\label{subsec:spectrum}

In the standard 1D KP model, energy spectrum exhibits a band structure, consisting of bands of allowed energies and energy gaps, where motion of a quantum particle is
forbidden. The size of energy gaps depends on the amplitude of delta interaction potential, and increases for stronger interactions. In this section we analyse how these
basic properties change, when one includes the transverse degrees of freedom, and when the particle-lattice node interaction has a finite range. By solving
Eq.~\eqref{eqn:SchroedU}, we obtain energy spectrum for an atom in quasi-1D geometry moving along infinite lattice of ions. This is compared to the analytical results of
a model assuming regularized contact pseudopotentials from Eq.~\eqref{final_eq1} for both constant and energy-dependent scattering length.

\begin{figure*}
		\includegraphics[width=0.32\textwidth]{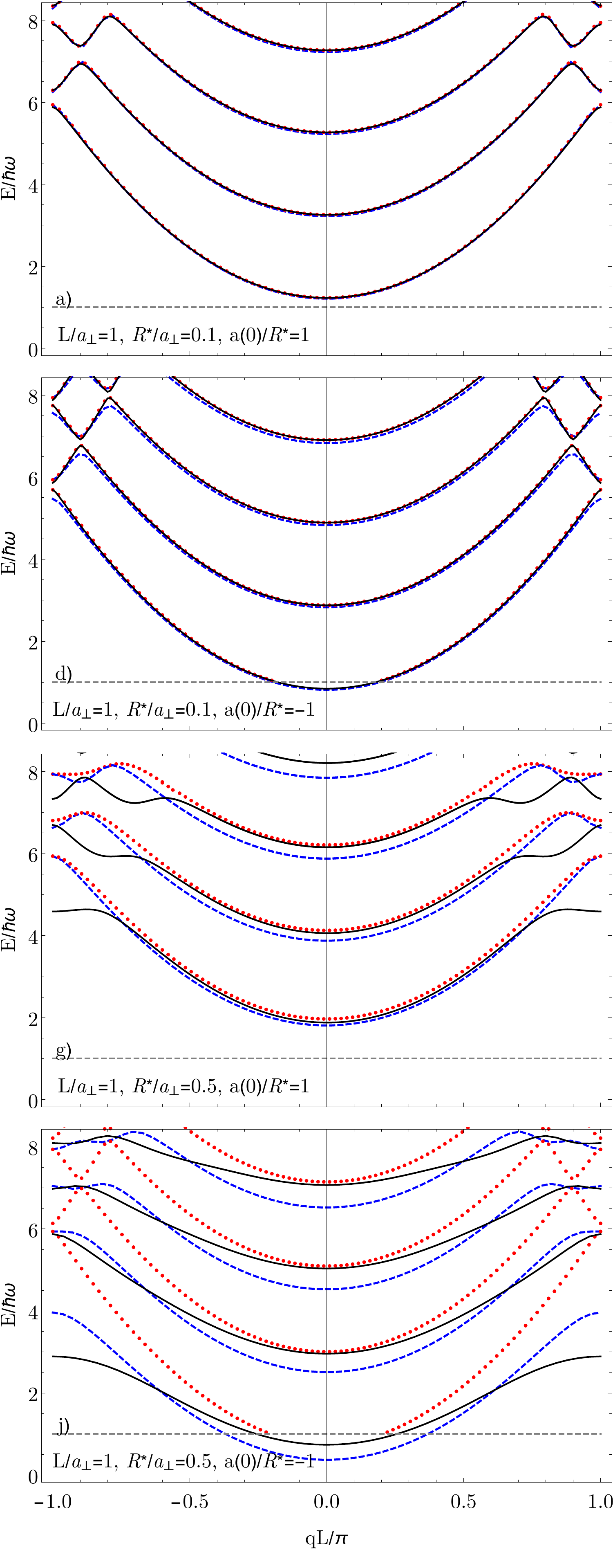}
		\includegraphics[width=0.32\textwidth]{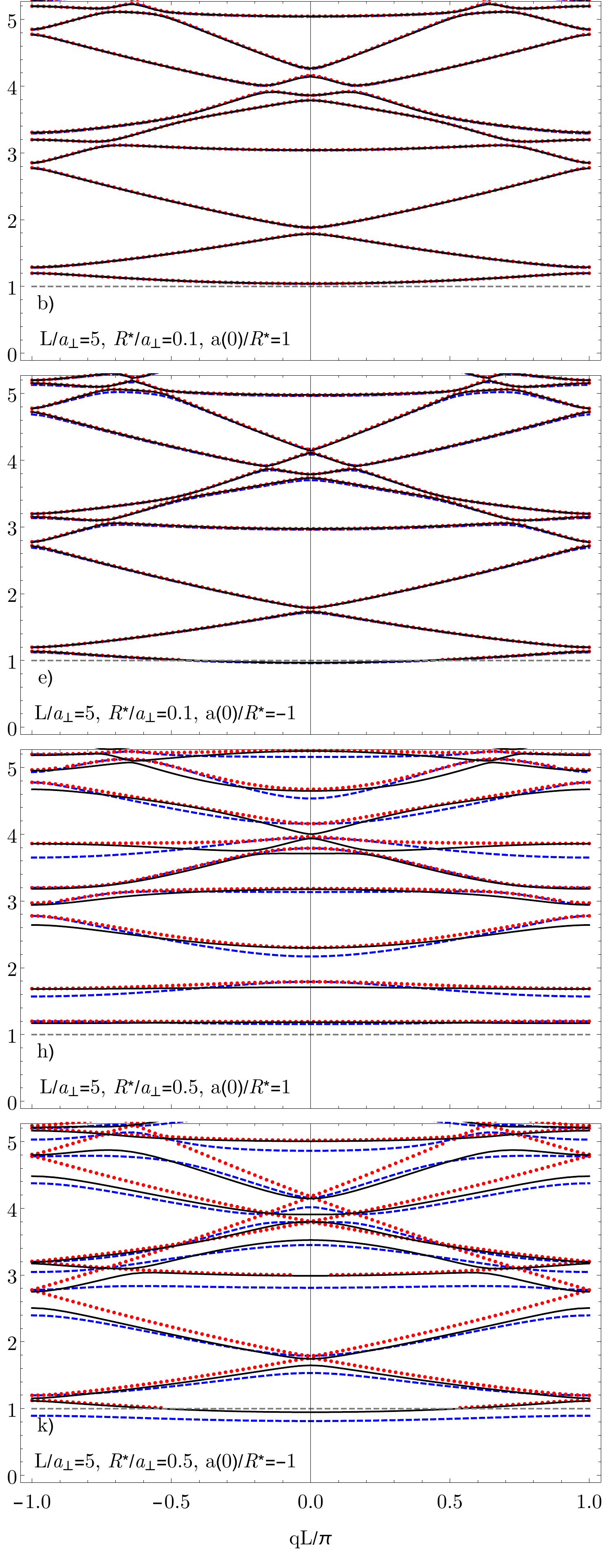}
		\includegraphics[width=0.32\textwidth]{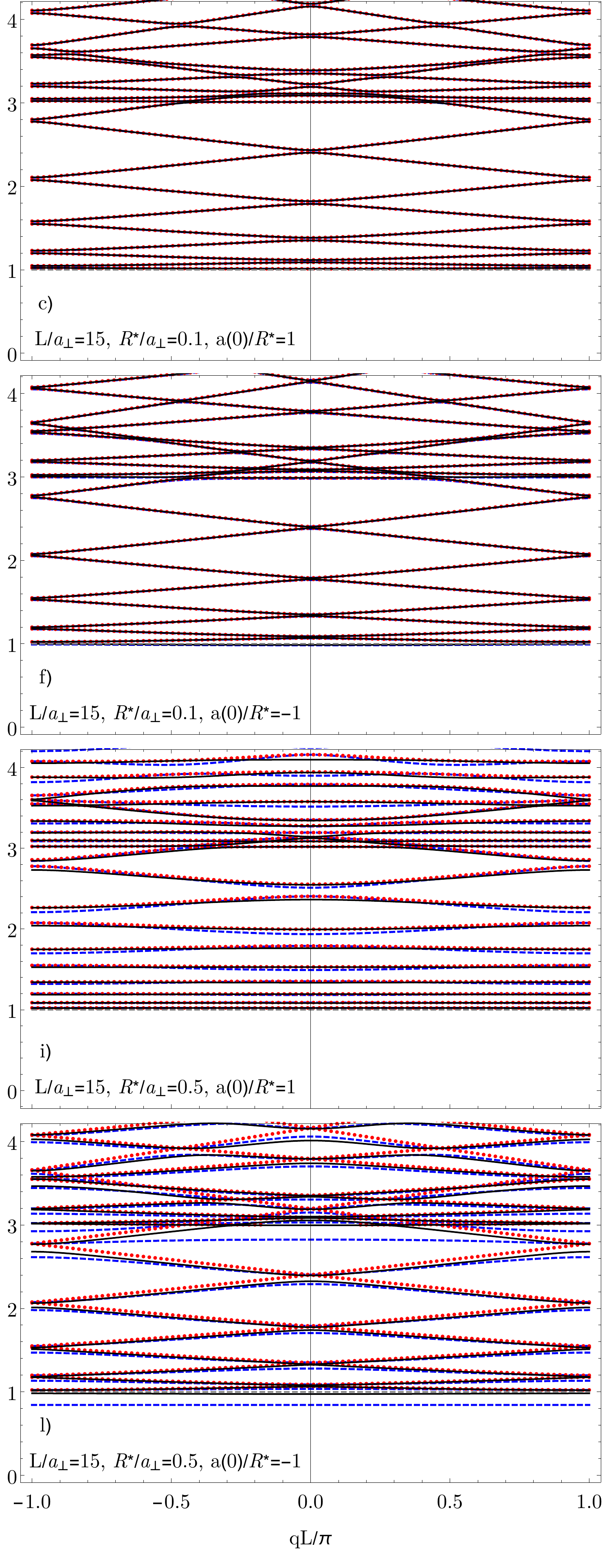}
	\caption{Energy spectrum of a quasi-1D KP model for the Fermi pseudopotential Eq.~\eqref{eqn:Fermi} (blue dashed line), 
	the pseudopotential interaction with energy-dependent scattering length \eqref{eqn:a1DOlshaniiE}) (red dotted line), and regularized atom-ion interaction \eqref{aeff} (black solid lines). The gray dashed line denotes $E = \hbar\omega$ as a reference.
    Calculations are performed for various lattice spacing $L/\aho$, the characteristic range of atom-ion potential $R^*/\aho$ and the scattering length $a/R^\ast$. The energy-dependent scattering length corresponds to atom-ion polarization potential with short-range boundary conditions parametrized with quantum-defect phase $\varphi$, see Eq.~\eqref{QuantDef}.}\label{fig:spectra}
\end{figure*}

Fig.~\ref{fig:spectra} shows the band structure calculated for different ratios of the distance $L$ between the ions, characteristic distance $R^*$, to the transverse
confinement length $a_{\perp}$. The atom-ion scattering length is expressed in units of $R^\ast$ as this quantity represents solely the properties of atom-ion interaction
potential, and it should not be scaled with size of external confinement. The range of parameters corresponds to current experiments on ultracold atom-ion systems, where
in order to minimize the effects of micromotion light atoms with heavy ions are preferably combined \cite{CetinaPRL12}. For instance, for $^6$Li atom interacting with
the linear chain of equally spaced $^{174}$Yb ions \cite{Joger2017} with $L=1.1\mu m$ confined in the transverse direction by harmonic trap with $\omega
~=~2~\pi~\times~100$ kHz, one obtains $L/\aho~=~15$ and $R^*/a_{\perp}~=~0.5$. For the same $L$ and $\omega$ parameters, in the case of Rb-$^{174}$Yb$^+$ system
\cite{Zipkes2010}, one gets $L/\aho~=~5$, $R^*/a_{\perp}~=~5$.

\begin{figure}
\includegraphics[width=0.5\textwidth]{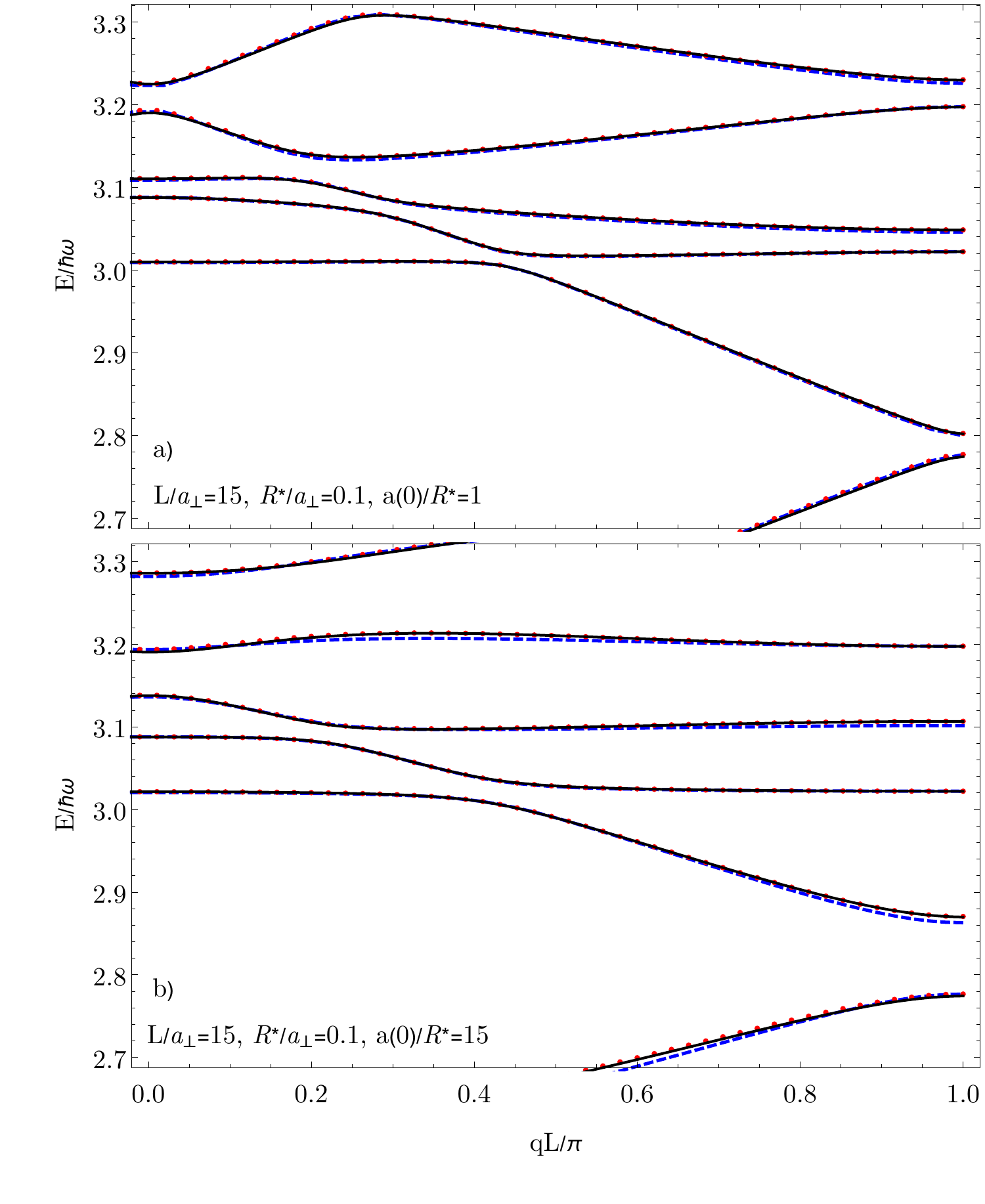}
\caption{Zoom in of the spectrum for $L/\aho=15$ with different values of $R^*/\aho$ and scattering length: (a)~$R^*/\aho=0.1$, $a/R^*=1$, (b)~ $R^*/\aho=0.1$, $a/R^*=15$.} \label{fig:spectrumL15b}
\end{figure}

Analysing Fig.~\ref{fig:spectra} we observe that for energies $E < 3 \hbar \omega$ the energy spectrum resembles the ordinary 1D KP model. This corresponds to the
situation with no excitations in the transverse direction, and the only effect of the transverse confinement is the renormalization of the 1D effective
scattering length \eqref{eqn:a1DOlshanii}. Above every threshold energy $E_{2n^*} = \hbar\omega(2 n^*(E) + 1)$, there is a new set of eigenstates appearing in the
spectrum, with 2D harmonic oscillator state in $n^\ast$-th excited state and $m = 0$ in the transverse direction. Each such a new set generates a new band structure, with
the lowest band starting at $E_{2n^*}$ and creating avoided crossings with band structures with smaller $n^\ast$. Fig.~\ref{fig:spectrumL15b} presents in more details
such avoided crosiings in the range of energies close to the second threshold at $3 \hbar \omega$. Another interesting feature is the size of the energy gaps for larger
ion separations: $L=5 R^\ast$ and $L=15^\ast$, which for $a=R^\ast$ is relatively large, as in the case of large coupling constants $g_\mathrm{1D}$ in 1D KP model. In
the other case $a=-R^\ast$, the energy gaps are relatively small and the spectrum is weakly affected in comparison to the dispersion relation of a free quantum particle
in quasi-1D geometry. This behaviour can be attributed to the renormalization of 1D scattering legth in quasi-1D confinement according to Eq.~\eqref{eqn:a1DOlshanii},
which for $a=R^\ast$ and $R^\ast = 0.5 a_\perp$ leads to much larger $g_\mathrm{1D}$ than for $a=-R^\ast$ and $R^\ast = 0.5 a_\perp$. In general, the numerically
calculated spectrum for relatively large ion separation $L$ and small rations $R^*/a_{\perp}$ are well approximated by models assuming pseudopotential interaction, in
particular when on includes the energy-dependent scattering length. The models assuming zero-range delta interactions break down when range of the potential $R^\ast$ becomes
comparable to the transverse confinement $a_\perp$, and for small ion separations $L$.

\begin{figure}
	\includegraphics[width=0.5\textwidth]{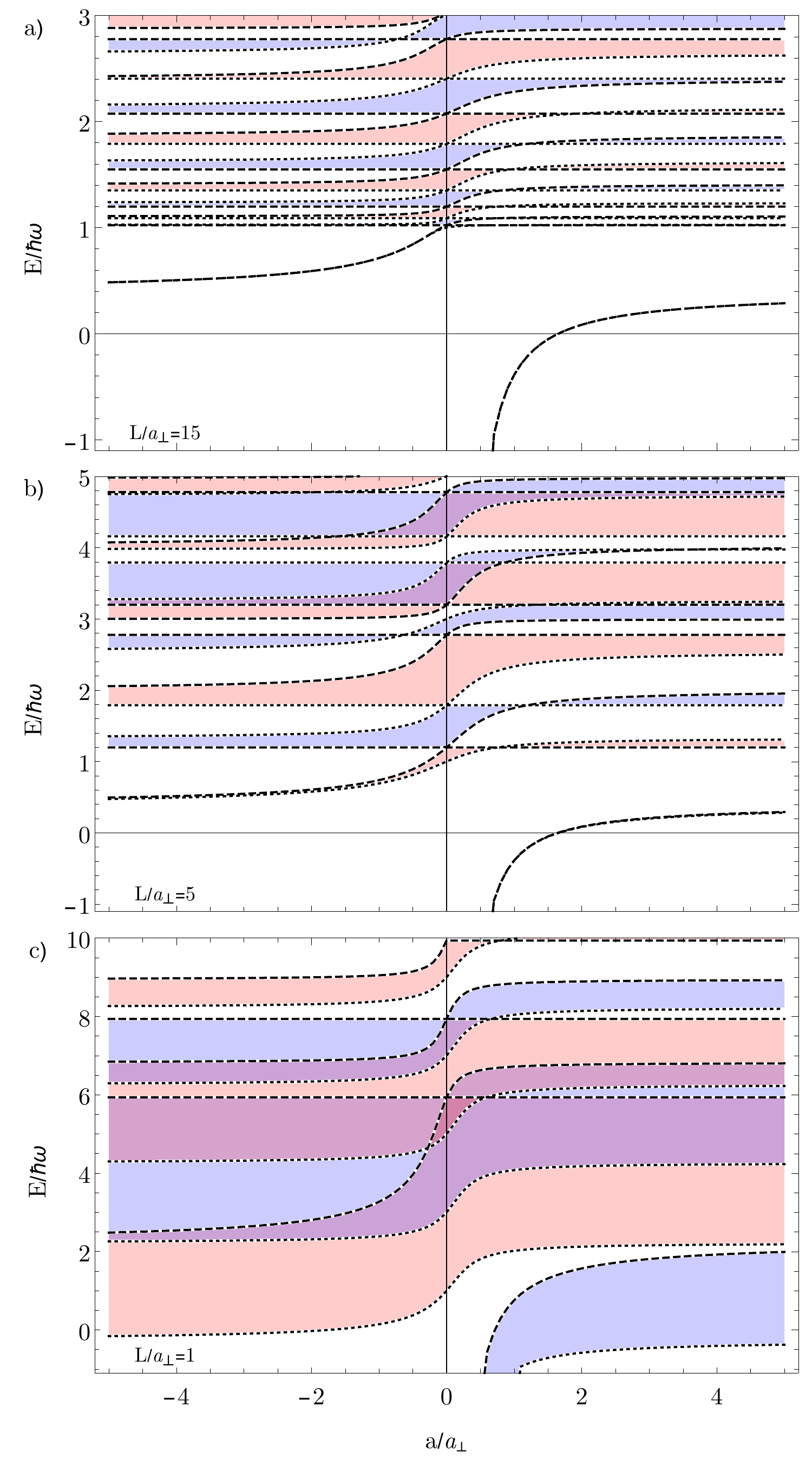}
	\caption{Energy bands as a function of the scattering length with (a) $L/a_{\perp}=15$, (b) $L/a_{\perp}=5$, (c) $L/a_{\perp}=1$. The black lines correspond to $qL/\pi=0$ (dotted) and $qL/\pi=1$ (dashed). The bands are colored in two different colors (pink and blue) for better readability.} 
	\label{fig:bandsA}
\end{figure}

Fig.~\ref{fig:bandsA} shows the dependence of the band structure (calculated for pseudopotential with energy-independent scattering length) on the scattering length. The black lines represent the boundaries of the spectrum: $qL/\pi=0$ (dotted) and $qL/\pi=1$ (dashed). The allowable values of energies for $qL/\pi\in (0,1)$ lie between these two lines and are marked with colors. 

In the case of large separation of the ions, bands are well separated and relatively narrow (Fig.~\ref{fig:bandsA}a). When the distance between the ions decreases, the bands  become more wide and start to overlap, which is visible especially in the case of $L/a_\perp=1$ (Fig.~\ref{fig:bandsA}c). It is worth noting that there are certain points where the bottom and upper boundaries of the band intersect, such as e.g. the second and third band in Fig.~\ref{fig:bandsA}b for positive scattering length. Another interesting feature is that the bound states ($E<\hbar\omega$) exist not only for positive scattering lengths, but also for $a<0$, similarly to bound states of two atoms confined in the harmonic traps \cite{Busch1998,Idziaszek2005}. 

\subsection{Wavefunctions}
\begin{figure*}
		\includegraphics[width=0.32\textwidth]{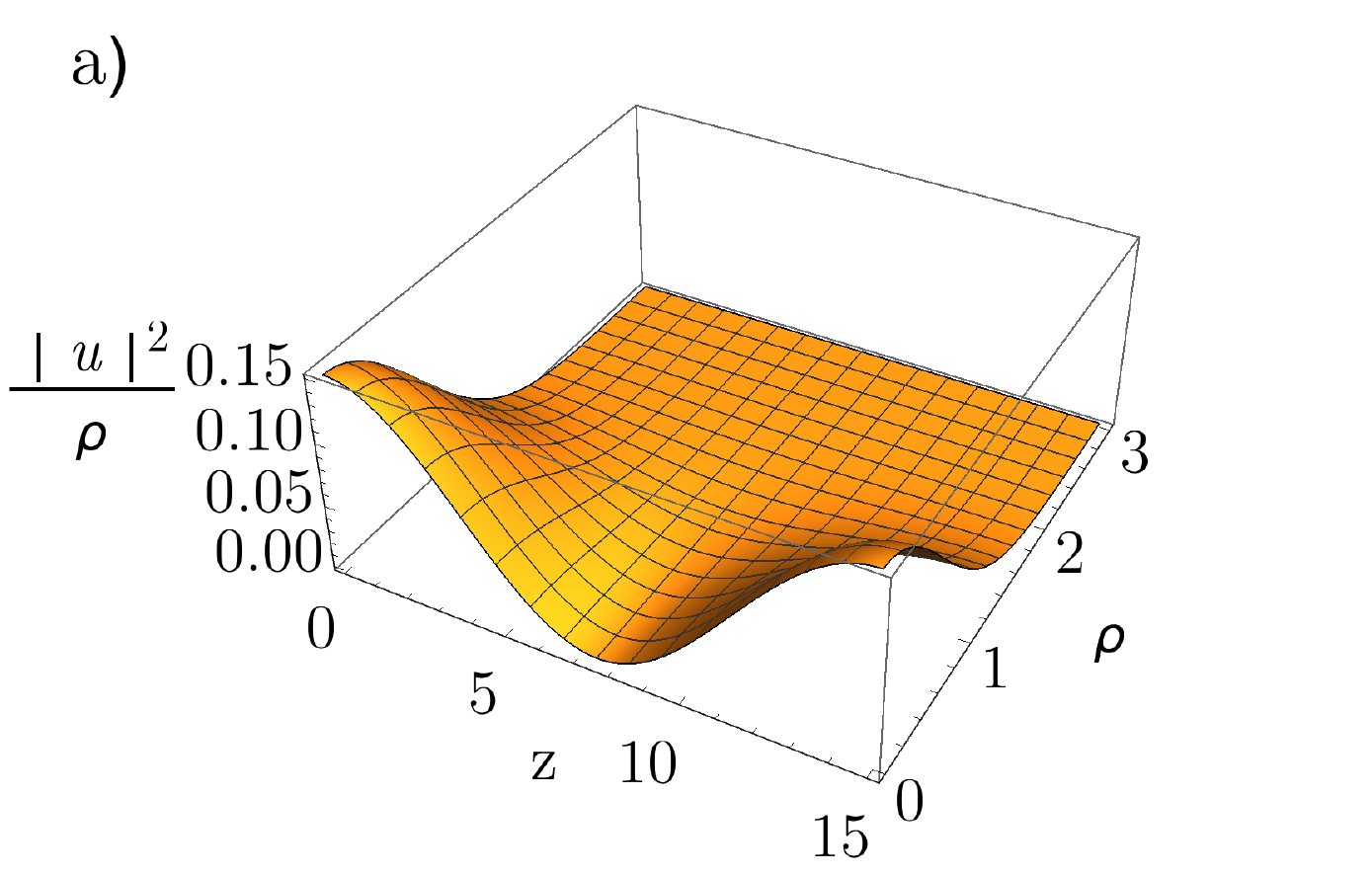}
		\includegraphics[width=0.32\textwidth]{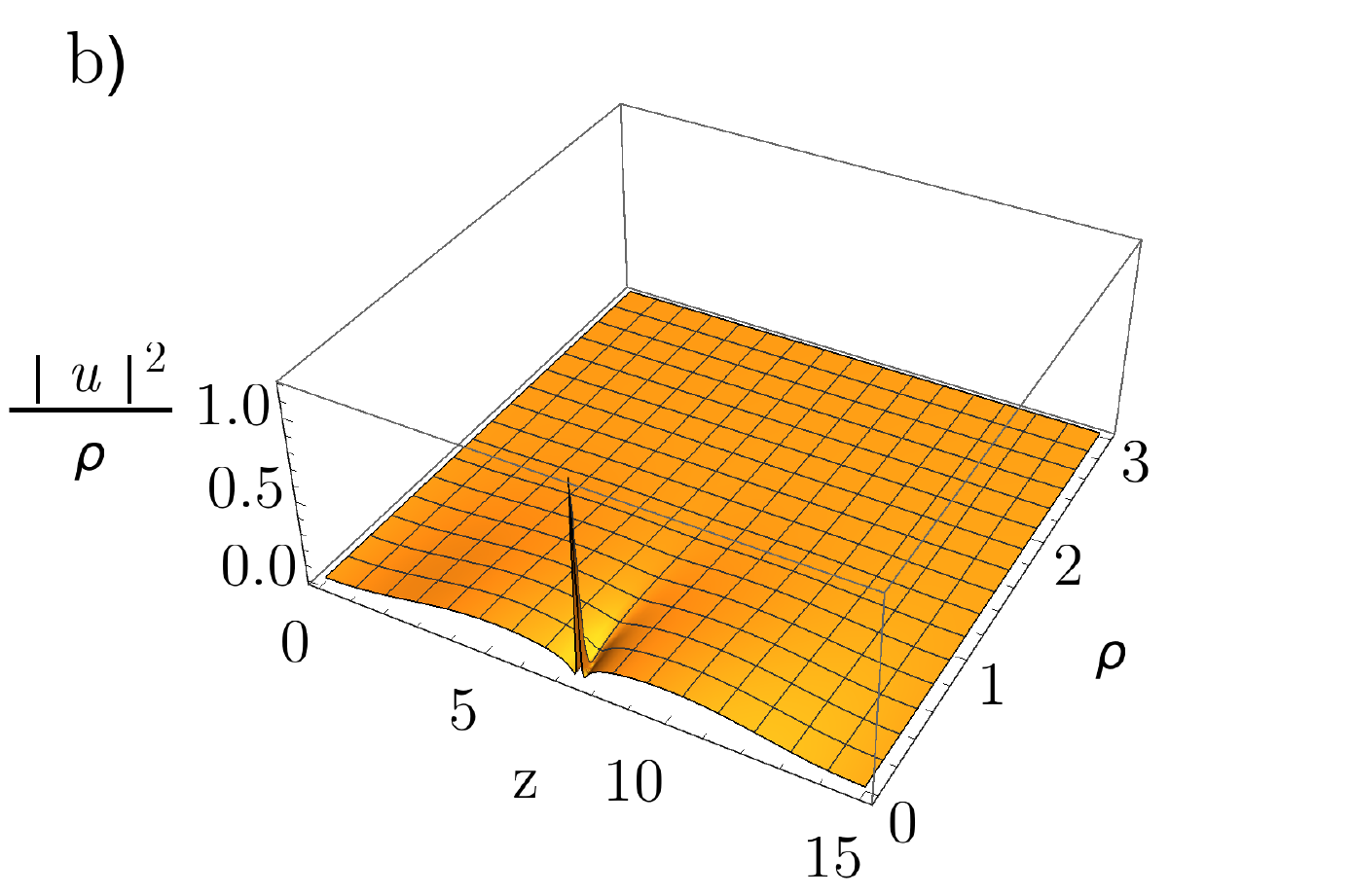}
		\includegraphics[width=0.32\textwidth]{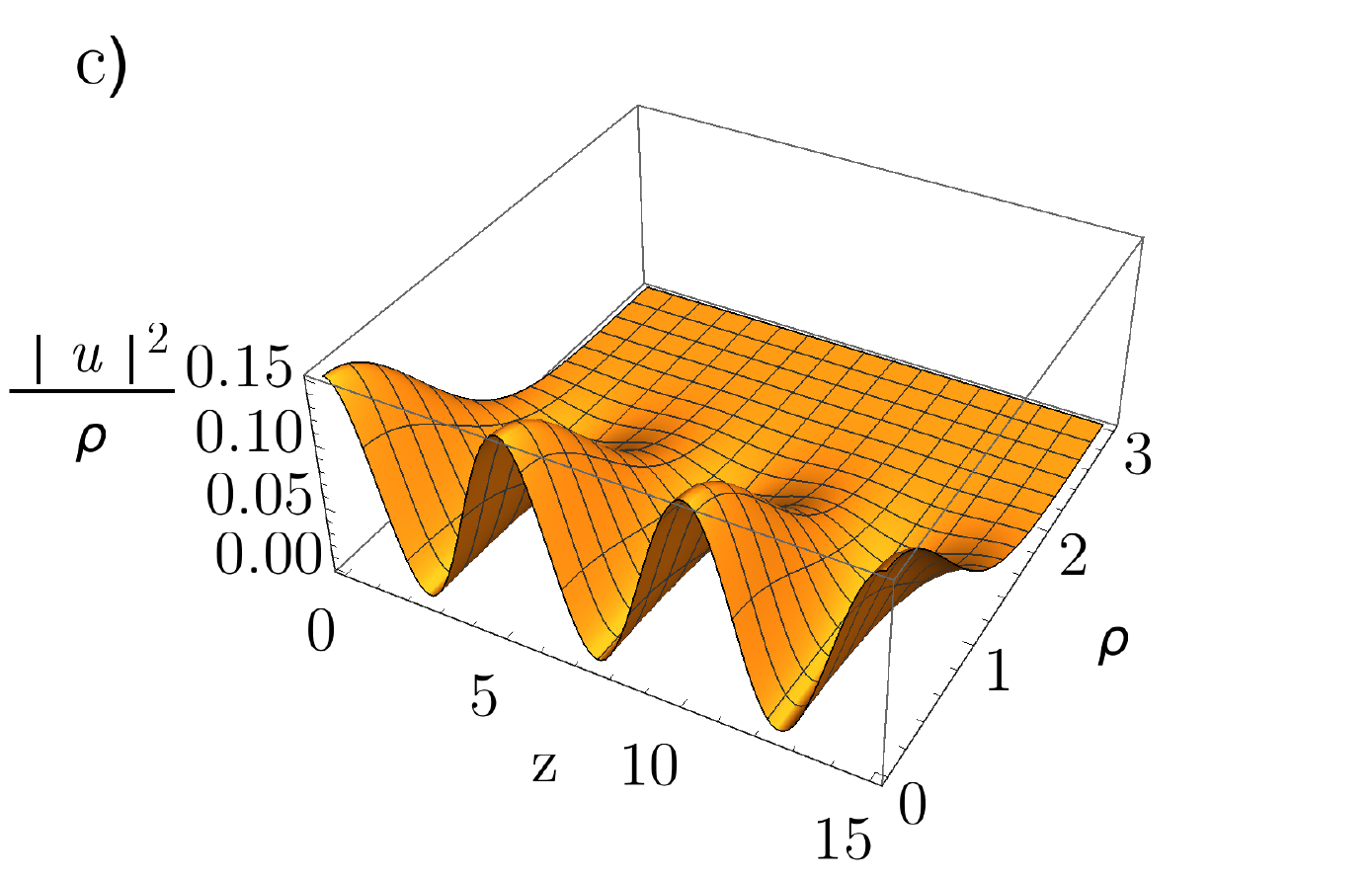}
		\includegraphics[width=0.32\textwidth]{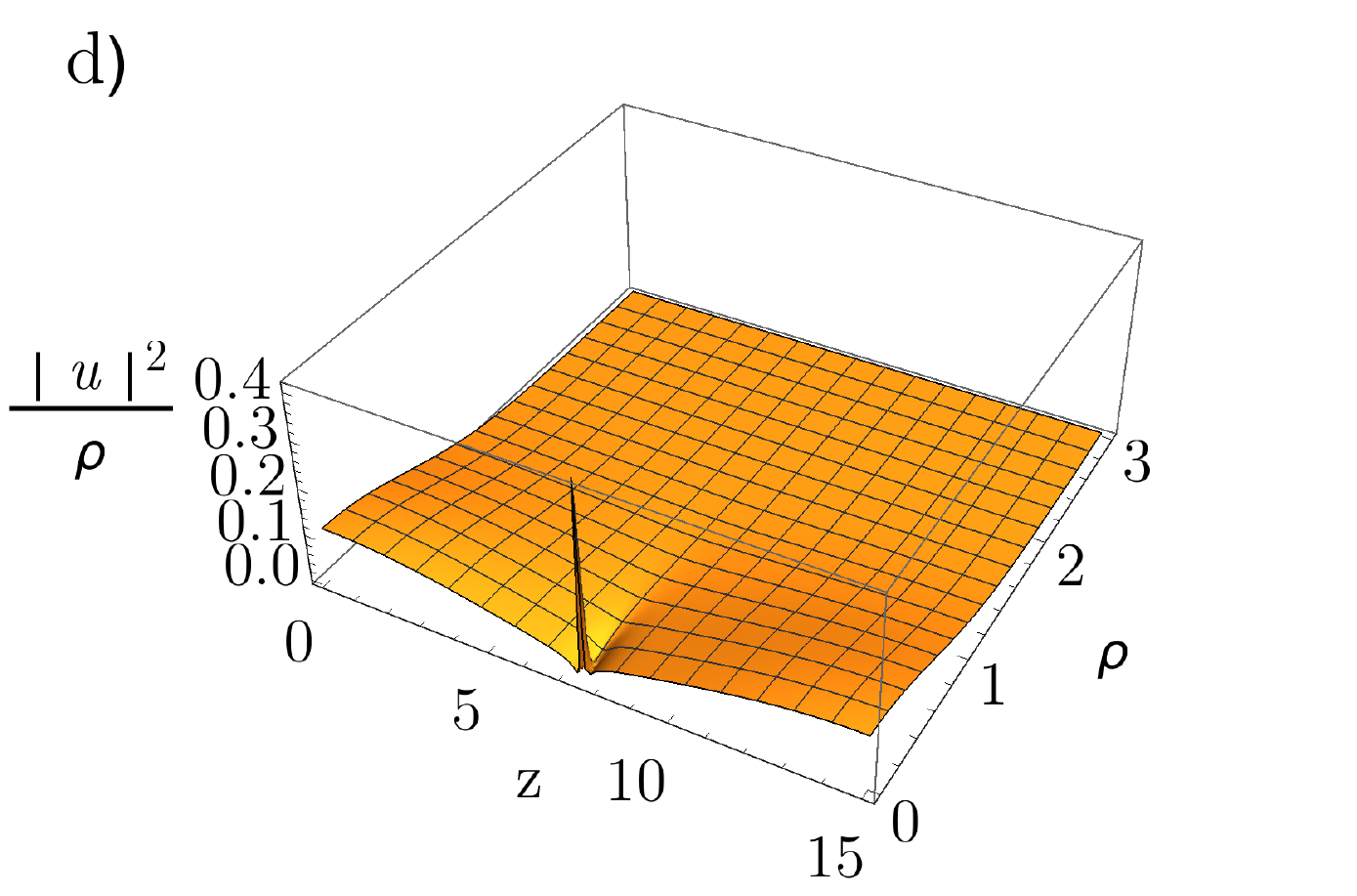}
		\includegraphics[width=0.32\textwidth]{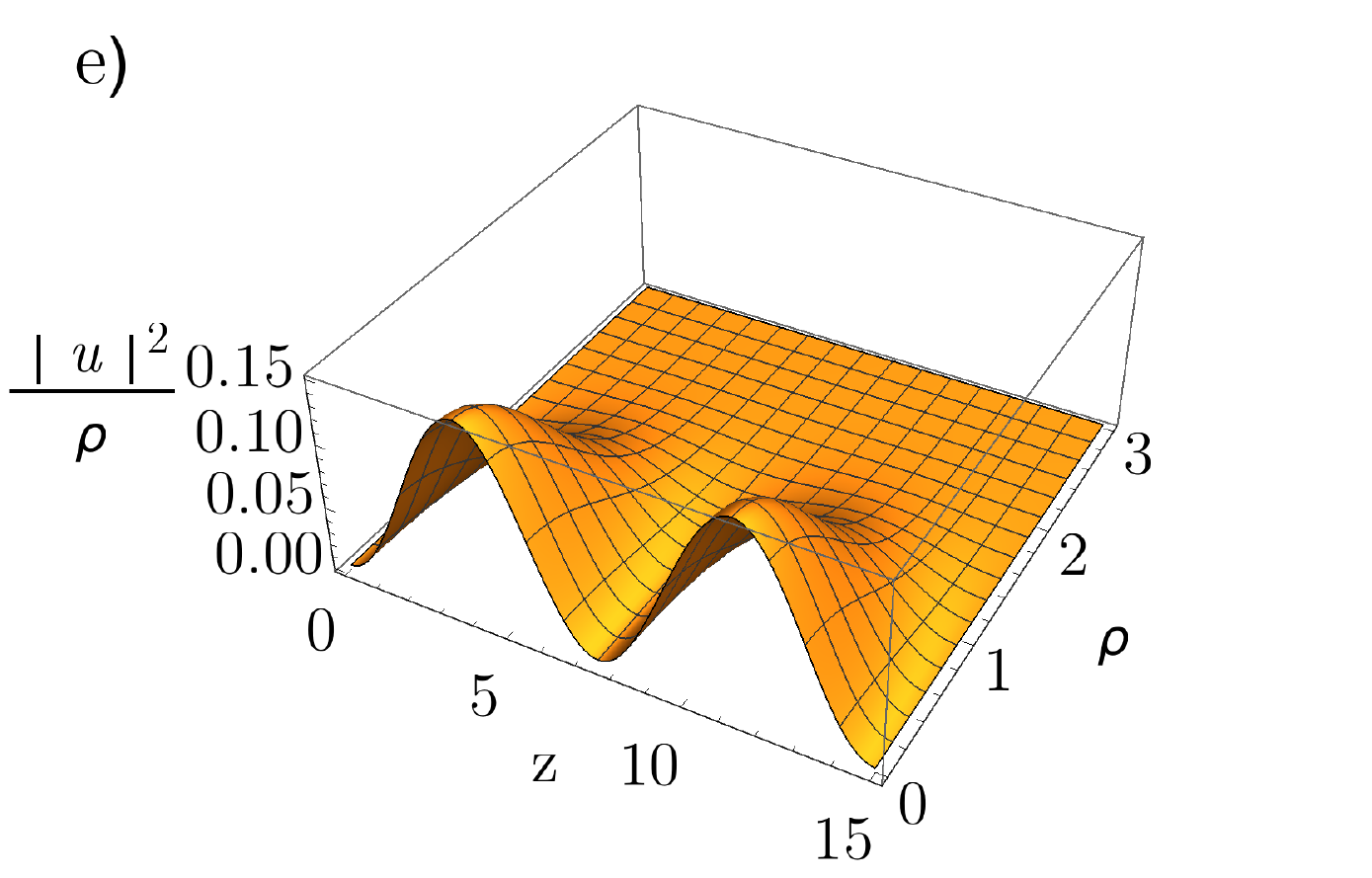}
		\includegraphics[width=0.32\textwidth]{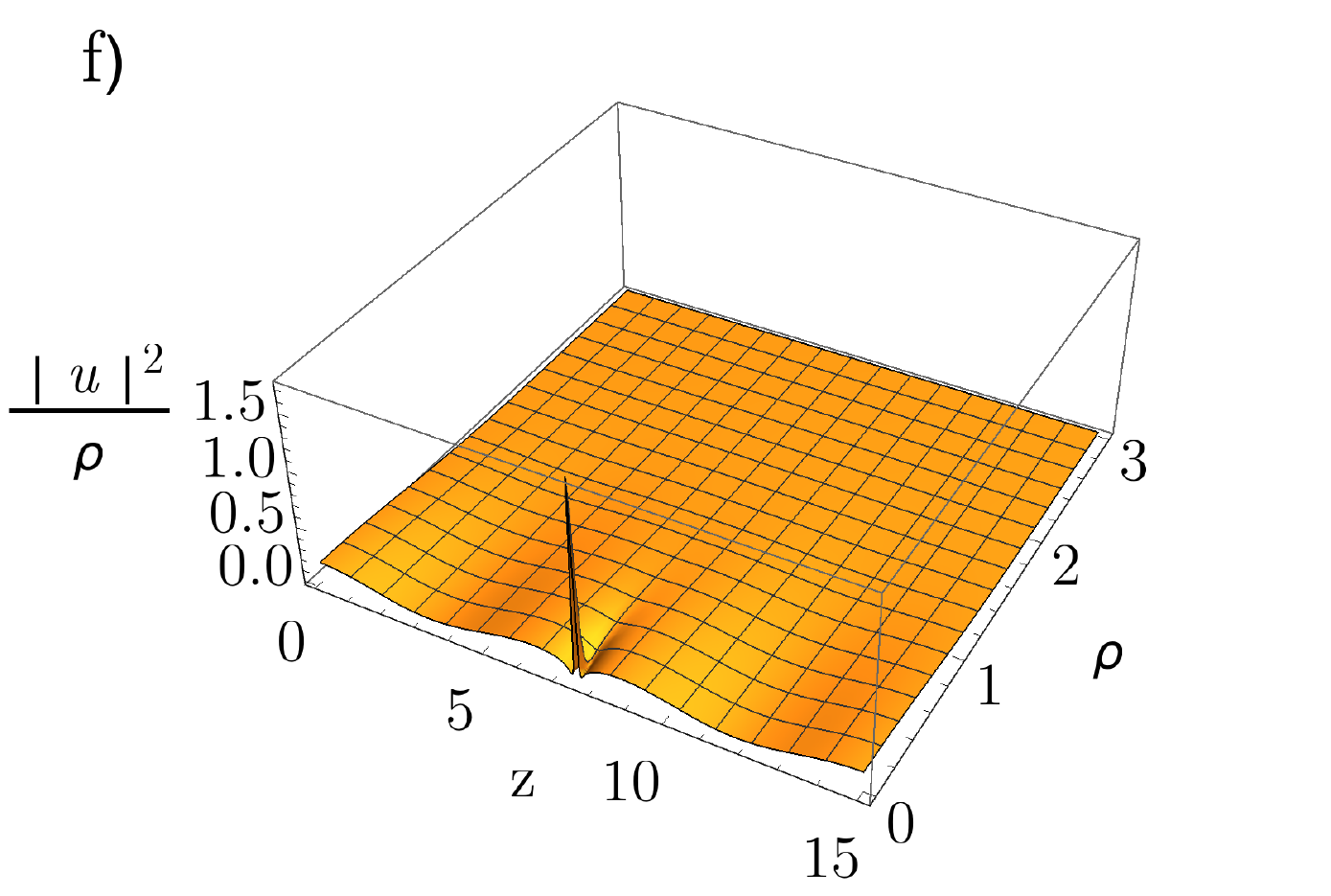}
	\caption{Wavefunctions calculated numerically within the model with the regularized atom-ion interaction potential: the lowest states of the spectrum for $L/\aho=15$, and $R^*/\aho=0.1$ and $a/R^*=1$. Upper panel (plots a-c) shows the wavefunctions for $q=-\pi/L$ and the lower panel (plots d-f) shows the states for $q=0$. Fig.~\ref{fig:wfpoints} shows the corresponding locations of the energy spectrum.} \label{fig:wf}
\end{figure*}

\begin{figure}
	\includegraphics[width=0.48\textwidth]{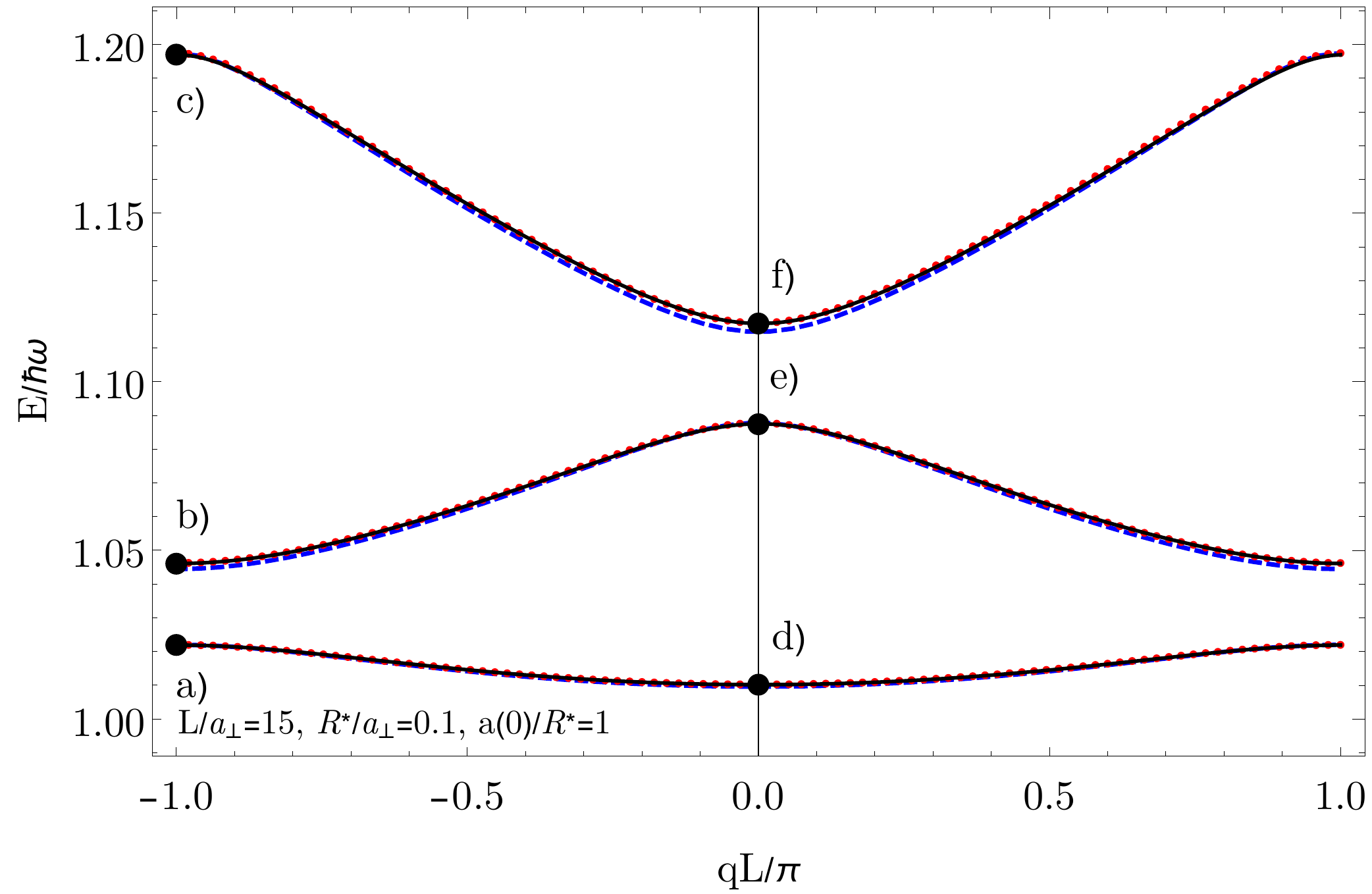}
	\caption{Three lowest states of the spectrum for $L/\aho=15$, $R^*/\aho=0.1$ and $a/R^*=1$. Black points refer to the parameters of Fig.  \ref{fig:wf}, where we plot the wavefunctions corresponding to each point on the spectrum.}\label{fig:wfpoints}
\end{figure}

Fig.~\ref{fig:wf} presents plots of the wavefunctions $|u_q(\rho, z)/\sqrt{\rho}|^2$ at $q=0$ and $q=-\pi/L$ for the three lowest states of the spectrum, evaluated numerically for regularized atom-ion potential \eqref{eqn:potRegB}, for $L/\aho=15$, $R^*/\aho=0.1$ and $a(0)=R^*$. 
Fig.~\ref{fig:wfpoints} shows the points on the energy spectrum corresponding to the wavefunctions plotted in Fig.~\ref{fig:wf}. It is worth noting that at the points (a), (c) and (e) the models assuming pseudopotential with constant and energy-dependent scattering length predict identical values. This is due to fact that at $z=L/2$ the wavefunction vanishes (see Fig.~\ref{fig:wf}a, c and e) exactly at the position, where the impurity is placed. Therefore, the presence of the impurity at this point does not affect the energy of the system, and the wavefunction are given by noninteracting particle states at this particular values of quasi-momenta. 

\subsection{Effective mass}

\begin{figure*}
	\includegraphics[width=\textwidth]{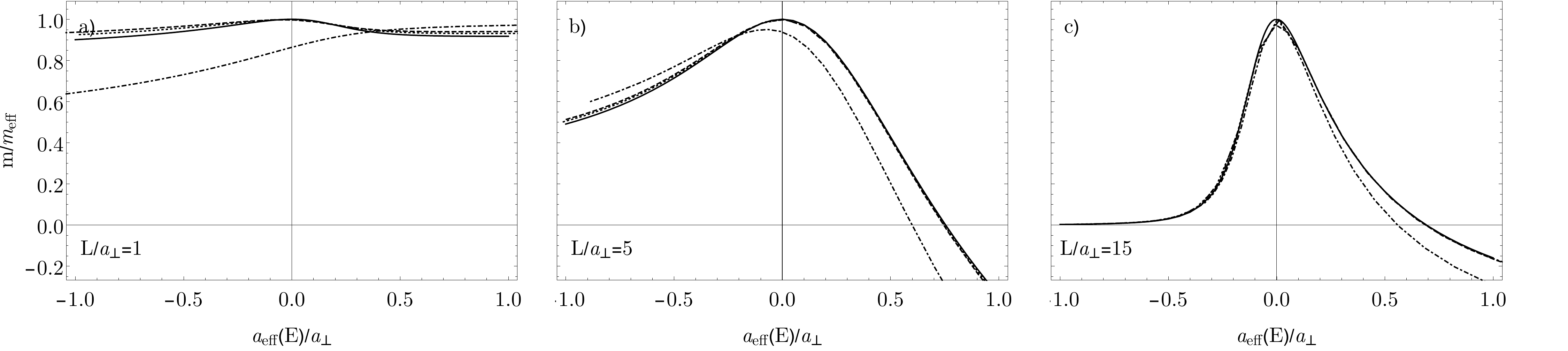}
	\caption{Effective mass for different values of $L/a_\perp$ evaluated for delta pseudopotential with constant scattering length (solid lines) and for regularized atom-ion potential, for different values of $R^*/a_{\perp}$: 0.05 (dotted), 0.1 (dashed), and 0.5 (dot-dashed). Panels shows the results for different ion's separations: (a) $L/\aho=1$, (b) $L/\aho=5$, (c) $L/\aho=15$} \label{fig:meff}
\end{figure*}

Effective mass is calculated by fitting the following dependence to the lowest band of the spectrum
\beq
E_0(q)=\epsilon_b + \frac{m}{m_\mathrm{eff}} |q|^2 + A|q|^4+B|q|^6,
\label{eqn:meff_fit}
\eeq
where  $\epsilon_b$ corresponds to the bottom of the energy band, and $A$ and $B$ are coefficients introduced in order to improve the quality of the fitting procedure at larger energies.  Eq.~\eqref{eqn:meff_fit} is expressed in dimensionless units of harmonic oscillator, and $m_{\mathrm{eff}}$ denotes the effective mass.
Fig.~\ref{fig:meff} shows how does this ratio change for different values of the scattering length. Both in the case of contact pseudopotential and regularized atom-ion potential, for large scattering length the effective mass becomes negative.
In the case of the regularized atom-ion potential, the energy-dependent scattering length (horizontal axis) is calculated numerically from Eq. \eqref{aeff} at the energy of the lowest band for $q=0$ ($\epsilon_b$ from Eq. \eqref{eqn:meff_fit}). The results for contact pseudopotential are calculated with constant scattering length.

The effective mass calculated within both models agrees best in case of small $R^*/a_{\perp}$ ratio, when the approximation of the atom-ion potential by contact interaction is applicable. The separation between the ions seems not crucial, however, for $L/a_{\perp}=1$ the results differ slightly in the region $|a|/a_{\perp} \sim 1$. 
Increasing the value of $R^*/a_{\perp}$ results in larger differences between both approaches, especially for positive scattering length $a$. At $a=0$ and in the limit $R^*/a \to 0$ the effective mass by definition is the same as physical mass $m_{\mathrm{eff}}=m$. At finite $R^*$, $m_{\mathrm{eff}}$ can be different than the physical mass even at $a=0$, which can be observed for the case of  $R^*/a_{\perp}=0.5$. This is due to the finite range effects of the atom-ion interaction potential.

\subsection{Effective scattering length for quasi-1D crystal}
\begin{figure}
		\includegraphics[width=0.5\textwidth]{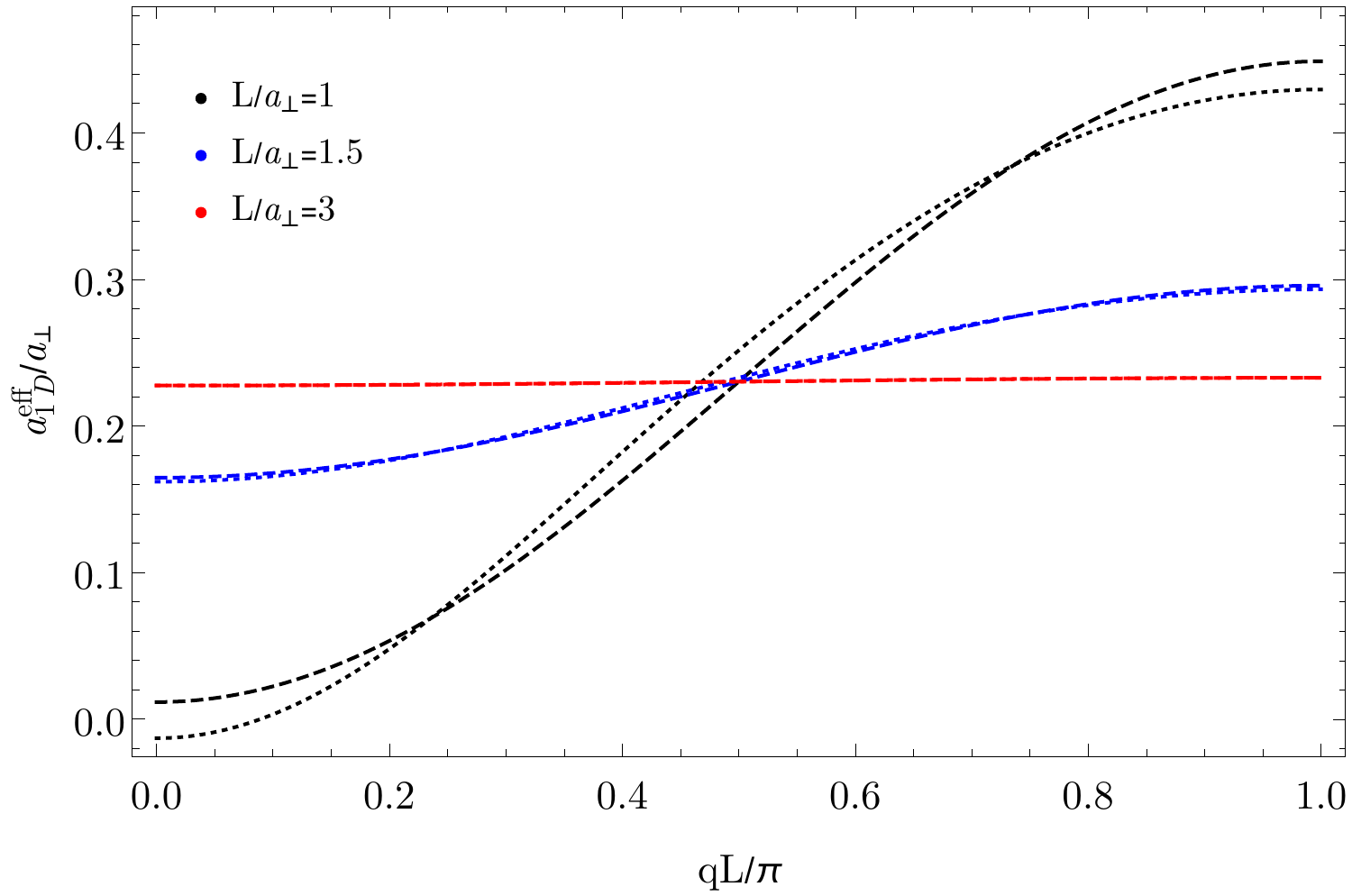}
	\caption{Effective one-dimensional scattering length $a_{1D}^{\mathrm{eff}}$ as defined in Eq.~\eqref{eqn:a1Deff} with $\Lambda_e$ given by Eq.~\eqref{eqn:LambdaEappr} (dotted lines) or Eq. \eqref{eqn:LambdaEapprWithH} (dashed lines) for $a/a_\perp=1$. Colors correspond to different values of $L/a_\perp$: $L/a_\perp=1$ (black lines), $L/a_\perp = 1.5$ (blue lines) and $L/a_\perp = 3$ (red lines).   } \label{fig:a1Deff}
\end{figure}

Fig.~\ref{fig:a1Deff} presents effective scattering length $a_{\mathrm{1D}}^{\mathrm{eff}}$ for a particle moving in quasi-1D crystal, determined from eq.~\eqref{eqn:a1Deff} with $\Lambda_e$ given by Eq.~\eqref{eqn:LambdaEappr} and Eq.~\eqref{eqn:LambdaEapprWithH}, at constant value of the scattering length. Calculation were performed for $a/a_{\perp}=1$ for three different values of $L/a_{\perp}$.
As expected, at large impurity separations $a_{\mathrm{1D}}^{\mathrm{eff}}$ does not depend on quasi-momentum, and its value corresponds to 1D scattering length of Olshanii \cite{Olshanii1998} for a single scattering centre. At smaller impurity separations $L=1.5 a_\perp$, $a_{\mathrm{1D}}^{\mathrm{eff}}$ starts to depend on the quasi-momentum. Finally at $L=a_\perp$, $a_{\mathrm{1D}}^{\mathrm{eff}}$ varies strongly with quasi-momentum, which indicates that scattering events on separate impurities interferes with each other, and one cannot describe the scattering in quasi-1D geometry using result \eqref{eqn:a1DOlshanii} derived for single scattering centre \cite{Olshanii1998}.
We note that for $L/a_{\perp}=1.5$  and $L/a_{\perp}=3$ approximation \eqref{eqn:LambdaEapprWithH} works relatively well, but for $L/a_{\perp}=1$ one has to sum up the series \eqref{eqn:LambdaEappr} in order to determine $\Lambda_e(E)$.

\section{Conclusions}
In our work we have generalized the Kronig-Penney model to quasi-1D geometry with harmonic confinement in the transverse direction. Such a system can be realized in experiments either by using tightly confined atoms of another species as affective lattice or using ionic crystal in hybrid atom-ion systems. Assuming regularized delta potential for the interaction, we have derived analytical equation determining energy spectrum as a function of quasi-momentum. In case of ionic crystal we have performed numerical calculations, that in the expected range of parameters agree well with the model assuming pseudopotential interaction. 

The main difference with respect to standard 1D KP model is the appearance of multiple overlapping bands, due to additional transverse degrees of freedom. In addition,
due to the transverse confinement the effective 1D scattering length exhibits resonances, allowing to reach the strong-coupling regime in KP model, where the energy
bands are separated by large gaps, already for atom-impurity scattering length comparable to the transverse confinement size. We have shown that is some regimes of
parameters, the effective mass determined at the bottom of the lowest energy band, becomes negative.

In the future studies, we plan to extend our model in order to incorporate the possibility of small oscillation of ions around equilibrium positions. In this approach we plan to include phonon excitations of the ionic crystal, and derive an effective model coupling atomic degrees of freedom with phonon excitations of the lattice. Such a setup can serve as a quantum simulator of the solid-state physics \cite{Bissbort2013}, and using presented results for delta pseudopotential interaction, it should be possible to determine analytically all the parameters of the effective Hamiltonian.

\section{Acknowledgement}
This work was supported by the Polish National Science Centre Grant No. 2015/17/B/ST2/00592. The authors would like to thank Krzysztof Jachymski for fruitful discussions.

\appendix

\section{1D Green's function $G_{1D}(z,z'|E)$}
\label{Sec:AppG1D}

The 1D Green's functions reads
\begin{equation}
\label{G1D}
G_{1D}^\pm(z,z'|E) = \frac{2m}{\hbar^2} \frac{1}{\pm 2 ik }e^{\pm i k |z-z'|},
\end{equation}
valid for $E>0$ with $k = \sqrt{2 m E/\hbar^2}$. For negative energy the two functions coincide and are exponentially decaying with the distance:
\begin{equation}
G_{1D}^\pm(z,z'|E) = \frac{2m}{\hbar^2} \frac{1}{- 2 \kappa}e^{-\kappa |z-z'|},
\end{equation}
with $\kappa= \sqrt{-2mE/\hbar^2}$ for $E<0$. Notice that it depends only on the distance between the points $z$ and $z'$.
It satisfies the following equation
\begin{equation}
\label{green_def}
\bigg(-\frac{\hbar^2}{2m} \frac{\partial^2}{\partial z^2} -  E \bigg)G_{1D}(z,z'|E) = - \delta(z-z').
\end{equation}
Formally, the Green's operator $\hat G(\zeta)$ is defined as the inverse of the operator $(\zeta-\frac{\hbar^2}{2m} \partial^2/\partial z^2)$ for arbitrary $\zeta$ for which 
the inverse exists. 
For $E<0$, the Green's function is defined as $G_{1D}^{+}(z,z'|E) = G_{1D}^{-}(z,z'|E) \equiv \bra{z} \hat G(E)\ket{z'}$, 
whereas $G_{1D}^{\pm}(z,z'|E) = \lim_{\epsilon\to0^+}\bra{z} \hat G(E\pm i \epsilon)\ket{z'}$.

\section{3D Green's function $G_{3D}^{\pm}(\x,\x'|E)$}
\label{G3Ddef}
The two 3D Green's functions $G_{3D}^+$ and $G_{3D}^-$, which describe outgoing and incoming waves at large distances, respectively, for the quasi-1D geometry are given by:
\begin{equation}
G_{3D}^{\pm}(\x,\x'|E) = \sum_{n,m} \phi_{n,m}(\xp)\phi_{n,m}^*(\xp')G_{1D}^{\pm}(z,z'|E-E_n),
\end{equation}
where the perpendicular variables are denoted by $\xp=(x,y)$, and the summation is taken of the radial quantum number of the 2D harmonic oscillator, $n=0,1,2,\ldots$, and
the azimuthal quantum number that, for fixed $n$, takes values $m=-n, -n+2, \ldots, n-2, n$. The eigenenergies of the 2D harmonic oscillator do not depend on the 
azimuthal quantum number and are given by $E_n = \hbar \omega (n+1)$.

The 2D harmonic oscillator functions are given by
\begin{eqnarray}
\phi_{n,m}(\xp) &=& \bigg[ \frac{2\alpha^2 p!}{(p+|m|)!} \bigg]^{1/2} e^{-\alpha^2 \rho^2/2} \times\\
&& \times(\alpha \rho)^{|m|} L_{p}^{|m|}(\alpha^2\rho^2)\frac{e^{im\varphi}}{\sqrt{2\pi}},
\end{eqnarray}
with harmonic oscillator length $\aho = \sqrt{\hbar/m\omega}$, $\alpha = 1/\aho$ and $p = (n-|m|)/2$.
Notice that at $\x=0$ the probability density is independent of the radial quantum number, i.e., $|\phi_{n,0}(0)|^2 = 1 / \pi \aho^2$.

\section{Evaluation of $\beta(E)$}
\label{Appbeta}
In this appendix, we demonstrate the necessary steps to evaluate the function $\beta(E)$ from Eq.~\eqref{betadef2}. Using the explicit form of the 3D Green's function, we can rewrite
\begin{equation}
\label{app-beta}
\beta(E) = \frac{\partial}{\partial z}\bigg[ z \sum_{n,m} |\phi_{2n,m}(\x_\perp = 0)|^2 G_{1D}^F(z| E - E_{2n})\bigg]\bigg|_{z=0^+},
\end{equation}
where the sum is over $n=0,1,2,\ldots$ and $m=0$ since $\phi_{\tilde n,m}(0)$ is zero for $|m|>0$ or odd $\tilde n$.
For completeness, we note that the Feynman Green's function is
\begin{equation}
\label{GF1D}
G_{1D}^F(z|E) = 
\left\{ \begin{array}{ll}
\frac{2m}{\hbar^2}  \frac{\sin k|z|}{2k} & \textrm{if $E\geqslant0$,}\\
\frac{2m}{\hbar^2} \frac{e^{- \kappa|z|}}{-2\kappa} & \textrm{if $E<0$,}\\
\end{array} \right.
\end{equation}
where $k = \sqrt{2m E/\hbar^2}$ for $E\geqslant0$ and $\kappa = \sqrt{2m (-E)/\hbar^2}$ for $E<0$. To proceed, for a fixed $E$, we set $n^*(E)$, so that the following
conditions are satisfied:
\begin{equation}
\label{nstar}
E_{2n^*} \leqslant E < E_{2n^*+2}.
\end{equation}
The energy $E$ can thus be written as $E = E_p + E_{2n^*}$, where $E_{2n^*}$ sets the threshold and $E_p$ is the available kinetic energy for the atom,
\begin{equation}
E_p = \frac{\hbar^2 p^2}{2m},
\end{equation}
which satisfies $0 \leqslant E_p < 2\hbar \omega$. Next, in Eq.~\eqref{app-beta}, we split the sum over $n$ into two contributions: for $n \leqslant n^*$ and $n>n^*$. In
the former, the argument of the Green's function $E-E_{2n}$ is non-negative, and so we use the upper formula from Eq.~\eqref{GF1D}. In the latter, the sum $E-E_{2n}$ is
negative, and so we apply the lower formula from Eq.~\eqref{GF1D}. We denote the sum with $n \leqslant n^*$ by $\beta_<$ and with $n>n^*$ by $\beta_>$; then $\beta =
\beta_< + \beta_>$. Explicit evaluation gives
\begin{equation}
\label{betaM}
\beta_<(E) = \frac{1}{\pi a_\perp^2} \frac{2m}{\hbar^2} \frac{\partial}{\partial z}\bigg[ z \sum_{0 \leqslant n \leqslant n^*}   \frac{\sin{p_n z}}{2 p_n}  \bigg]\bigg|_{z=0^+},
\end{equation}
where $z>0$ and $p_n = \sqrt{2m |E- E_{2n}|/\hbar^2}$. The remaining part is given by
\begin{equation}
\label{betaP}
\beta_>(E) = \frac{1}{\pi a_\perp^2} \frac{2m}{\hbar^2} \frac{\partial}{\partial z}\bigg[ z \sum_{n > n^*}   \frac{e^{-p_n z}}{-2 p_n}  \bigg]\bigg|_{z=0^+}.
\end{equation}
There is always a finite sum in $\beta_<$, and so the derivative can be evaluated explicitly term by term. Due to the condition $z=0^+$, we have $\beta<(E) = 0$, and, therefore,
$\beta(E) = \beta_>(E)$.

Now, we rewrite Eq.~\eqref{betaP} in the following form:
\begin{equation}
\beta(E) = -\frac{a_\perp}{2} \frac{1}{\pi a_\perp^2} \frac{2m}{\hbar^2} \frac{\partial}{\partial x}\bigg[ x \sum_{n > n^*} \frac{e^{ -\alpha_n x}}{\alpha_n}
\bigg]\bigg|_{x=0^+},
\end{equation}
where we introduced $x = z/a_\perp$ and $\alpha_n = \sqrt{4n - 2 (E-\hbar\omega)/\hbar\omega}$. We can further simplify the sum by shifting the index $n \to n + n^*$. Then, we obtain
\begin{equation}
\beta(E) = -\frac{a_\perp}{2} \frac{1}{\pi a_\perp^2} \frac{2m}{\hbar^2} \frac{\partial}{\partial x}\bigg[ x \sum_{n=1}^\infty \frac{e^{ -\tilde\alpha_n x}}{\tilde\alpha_n}
\bigg]\bigg|_{x=0^+},  
\end{equation}
where $\alpha_n = \sqrt{4n - 2 \epsilon}$ and $\epsilon = (E - E_{2n^*})/\hbar\omega$ with $0 \leqslant \epsilon <2$. In this form, the sum can be evaluated explicitly and reads:
\begin{equation}
\frac{\partial}{\partial x}\bigg[ x \sum_{n=1}^\infty \frac{e^{ -\tilde\alpha_n x}}{\tilde\alpha_n}
\bigg]\bigg|_{x=0^+} = \frac{1}{2} \zeta_H \bigg(\frac{1}{2}, 1 - \frac{\epsilon}{2}\bigg),
\end{equation}
where $\zeta_H$ is the Hurwitz zeta function. Finally, we obtain $\beta(E)$ given in Eq.~\eqref{beta_final}.

\section{Evaluation of $\Lambda(E) = \Lambda_p(E) + \Lambda_e(E)$}
\label{AppLambda}
In this appendix, we demonstrate the evaluation of $\Lambda(E)$ from Eq.~\eqref{defLambda}, i.e.,
\begin{equation}
\label{app-lambda}
\Lambda(E) = \sum_{M>0}  2  G_{3D}( M L |E)\, \cos{M\theta},
\end{equation}
where the sum is over integer positive $M$.
The explicit form of the 3D Green's function reads:
\begin{eqnarray}
G_{3D}(ML|E) &=& \frac{1}{\pi a_\perp^2} \bigg[ \sum_{0 \leqslant n\leqslant n^*}    G_{1D}^F( M L |E-E_{2n}) + \\
&& + \sum_{n>n^*}    G_{1D}^F( M L |E-E_{2n}) \bigg],
\end{eqnarray}
where we used the definition of $n^*(E)$, see Eq.~\eqref{nstar}. Now, in the first term, $E-E_{2n}$ is non-negative, and so we apply the upper formula from Eq.~\eqref{GF1D}.
In the second term, $E-E_{2n}$ is negative, and here we apply the lower formula from Eq.~\eqref{GF1D}. In this way, we arrive at
\begin{eqnarray}
G_{3D}(ML|E) &=& \frac{1}{\pi a_\perp^2} \frac{2m}{\hbar^2} \bigg[ \sum_{ 0 \leqslant n\leqslant n^*}    \frac{\sin( M p_n L )}{2 p_n}  + \nonumber\\
&& + \sum_{n>n^*}    \frac{e^{-M p_n L }}{-2 p_n} \bigg],
\label{app-g3d}
\end{eqnarray}
where $p_n = \sqrt{2m |E-E_{2n}|/\hbar^2}$. 
Now, we insert the form of the Green's function from Eq.~\eqref{app-g3d} into Eq.~\eqref{app-lambda}, and write:
\begin{equation}
\Lambda(E) = \frac{L}{\pi a_\perp^2} \frac{2m}{\hbar^2}[ \Lambda_p(E) + \Lambda_e(E) ],
\end{equation}
where
\begin{subequations}
	\begin{eqnarray}
	\Lambda_p(E) &=& \sum_{M>0 }\sum_{ n=0}^{n^*}    \frac{2\cos(M\theta)\sin( M p_n L )}{2 L p_n} \\
	\Lambda_e(E) &=& \sum_{M>0} \sum_{n = n^*+1}^\infty    \frac{2\cos(M\theta) e^{-M p_n L }}{-2 L p_n}.
	\end{eqnarray}
\end{subequations}
In this expressions, the sums over $M$ can be explicitly evaluated. As a result, we obtain for $\Lambda_p(E)$ and $\Lambda_e(E)$ Eqs.~\eqref{eqn:LambdaP}
and~\eqref{eqn:LambdaE}, respectively.


\bibliography{bibliografia}
\bibliographystyle{apsrev4-1}
\end{document}